\documentclass[aps,prb,reprint,superscriptaddress,citeautoscript,floatfix,longbibliography,bibnotes]{revtex4-2}
\usepackage{times,siunitx,amsmath,amsfonts,amsthm,amssymb,amsbsy,braket,graphicx,orcidlink}
\usepackage[utf8]{inputenc}
\usepackage[T1]{fontenc}
\usepackage[capitalize]{cleveref}
\newcommand{\ii}{\mathrm{i}}
\newcommand{\ee}{\mathrm{e}}
\newcommand{\dd}{\mathrm{d}}
\newcommand{\up}{{\uparrow}}
\newcommand{\down}{{\downarrow}}
\renewcommand{\Re}{\mathrm{Re}}
\renewcommand{\Im}{\mathrm{Im}}

\DeclareMathOperator{\sgn}{sgn}
\DeclareMathOperator{\Tr}{Tr}
\DeclareMathOperator{\pf}{pf}
\DeclareMathOperator{\diag}{diag}
\hypersetup{colorlinks=true, urlcolor=blue, linkcolor=blue, citecolor=red}

\begin{document}
\title{
Dispersive 1D Majorana modes with emergent supersymmetry in 1D proximitized superconductors via spatially-modulated potentials and magnetic fields
}
\author{Pasquale Marra \orcidlink{0000-0002-9545-3314}}
\email{pmarra@ms.u-tokyo.ac.jp}
\affiliation{
Graduate School of Mathematical Sciences,
The University of Tokyo, 3-8-1 Komaba, Meguro, Tokyo, 153-8914, Japan}
\affiliation{
Department of Physics, and Research and Education Center for Natural Sciences, 
Keio University, 4-1-1 Hiyoshi, Yokohama, Kanagawa, 223-8521, Japan}
\author{Daisuke Inotani \orcidlink{0000-0002-7300-1587}}
\author{Muneto Nitta \orcidlink{0000-0002-3851-9305}}
\affiliation{
Department of Physics, and Research and Education Center for Natural Sciences, 
Keio University, 4-1-1 Hiyoshi, Yokohama, Kanagawa, 223-8521, Japan}
\date{\today}

\begin{abstract}
In condensed matter systems, zero-dimensional or one-dimensional Majorana modes can be realized respectively as the end and edge states of one-dimensional and two-dimensional topological superconductors.
In this \emph{top-down} approach, $(d-1)$-dimensional Majorana modes are obtained as the boundary states of a topologically nontrivial $d$-dimensional bulk.
In a \emph{bottom-up} approach instead, $d$-dimensional Majorana modes in a $d$-dimensional system can be realized as the continuous limit of a periodic lattice of coupled $(d-1)$-dimensional Majorana modes.
We illustrate this idea by considering one-dimensional proximitized superconductors with spatially-modulated potential or magnetic fields.
The ensuing inhomogeneous topological state exhibits one-dimensional counterpropagating Majorana modes with finite dispersion, and with a Majorana gap that can be controlled by external fields.
In the massless case, the Majorana modes have opposite Majorana polarizations and pseudospins, are conformally invariant, and realize centrally extended quantum mechanical supersymmetry.
The supersymmetry exhibits spontaneous partial breaking.
Consequently, the massless Majorana fermion can be identified as a Goldstino, i.e., the Nambu-Goldstone fermion associated with the spontaneously broken supersymmetry.
\end{abstract}
\maketitle

\section{Introduction}

In condensed matter, zero-dimensional (0D) Majorana modes~\cite{kitaev_unpaired_2001,oreg_helical_2010,lutchyn_majorana_2010} can emerge as quasiparticle excitations corresponding to topologically protected end states in the nontrivial phase of 1D superconductors~\cite{qi_topological_2011,leijnse_introduction_2012,stanescu_majorana_2013,beenakker_majorana_2013,aguado_majorana_2017,sato_topological_2017,shen_topological_2011}, realized in proximitized nanowires~\cite{stanescu_majorana_2013,mourik_signatures_2012,lee_zerobias_2012,rokhinson_fractional_2012,das_zerobias_2012,deng_anomalous_2012,finck_anomalous_2013,churchill_superconductor_2013,lee_spinresolved_2014,deng_majorana_2016,nichele_scaling_2017,chen_experimental_2017,gul_ballistic_2018,grivnin_concomitant_2019} or quantum chains of adatoms~\cite{choy_majorana_2011,pientka_topological_2013,nadj-perge_observation_2014,pawlak_probing_2016,feldman_highresolution_2017,kim_tailoring_2018,pawlak_majorana_2019} deposited on top of a superconductor.
Moreover, chiral and helical 1D Majorana modes have been predicted to emerge as topologically protected edge states in the topologically nontrivial phase of a 2D superconductor or planar superconducting heterostructures~\cite{qi_time-reversal-invariant_2009,qi_chiral_2010,nakosai_topological_2012,seradjeh_majorana_2012,zhang_timereversalinvariant_2013,sun_helical_2016,he_chiral_2017,chen_helical_2018,he_platform_2019,li_majorana-josephson_2019,hu_chiral_2019,hogl_chiral_2020,kayyalha_absence_2020,shen_spectroscopic_2020,he_optical_2021}.
In chiral topological superconductors, each edge carries one mode with a definite chirality, whereas in helical topological superconductors, each edge carries a helical pair of counterpropagating modes having opposite spins.
In any case, 0D and 1D (chiral or helical) Majorana modes correspond to the boundary excitations of a topologically nontrivial bulk~\cite{qi_time-reversal-invariant_2009,qi_chiral_2010,nakosai_topological_2012,seradjeh_majorana_2012,zhang_timereversalinvariant_2013,sun_helical_2016,he_chiral_2017,chen_helical_2018,he_platform_2019,li_majorana-josephson_2019,hu_chiral_2019,hogl_chiral_2020,kayyalha_absence_2020,shen_spectroscopic_2020,he_optical_2021}.

Despite the fact that theoretical models are rather simple, the experimental quest for signatures of Majorana quasiparticles has been challenging~\cite{stanescu_majorana_2013,mourik_signatures_2012,lee_zerobias_2012,rokhinson_fractional_2012,das_zerobias_2012,deng_anomalous_2012,finck_anomalous_2013,churchill_superconductor_2013,lee_spinresolved_2014,deng_majorana_2016,nichele_scaling_2017,chen_experimental_2017,gul_ballistic_2018,grivnin_concomitant_2019,choy_majorana_2011,pientka_topological_2013,nadj-perge_observation_2014,pawlak_probing_2016,feldman_highresolution_2017,kim_tailoring_2018,pawlak_majorana_2019,he_chiral_2017,kayyalha_absence_2020,shen_spectroscopic_2020} since identical signatures can be explained without invoking Majorana quasiparticles in the presence of inhomogeneous potentials and disorder~\cite{chevallier_mutation_2012,liu_zerobias_2012,kells_nearzeroenergy_2012,stanescu_disentangling_2013,roy_topologically_2013,san-jose_majorana_2016,liu_andreev_2017,liu_distinguishing_2018,moore_quantized_2018,ricco_spindependent_2018,chen_ubiquitous_2019,pan_generic_2019,prada_andreev_2020}.
However, the experimental effort is justified by the fact that they could provide the building blocks for topological quantum computers due to the non-abelian nature of their braiding statistics~\cite{ivanov_nonabelian_2001,kitaev_faulttolerant_2003,alicea_nonabelian_2011,sarma_majorana_2015,aasen_milestones_2016,karzig_scalable_2017,lian_topological_2018}.
This could lead to a major technological advance in the field of quantum computing in the not too distant future.
Besides, Majorana modes are intrinsically interesting from a purely theoretical perspective because they could provide the platform to study fundamental physics theories.
Specifically, supersymmetry (SUSY) is a fundamental symmetry between bosons and fermions that has been introduced to solve the hierarchy problem in high-energy physics, which extends the Standard Model introducing a superpartner to each known elementary particle~\cite{witten_dynamical_1981}.
Generally, quantum mechanical SUSY is any symmetry that relates many-body states with opposite fermion parity~\cite{cooper_supersymmetry_1995,gangopadhyaya_supersymmetric_2017}.
It has been shown that chains of 0D Majorana modes exhibit emergent SUSY~\cite{hsieh_all_2016,huang_supersymmetry_2017,rahmani_emergent_2015,rahmani_phase_2015,rahmani_interacting_2019} and that the time-reversal symmetry which protects helical 1D Majorana modes can be identified with quantum mechanical SUSY~\cite{qi_time-reversal-invariant_2009}.

This work aims to theoretically describe in a general framework the realization of dispersive Majorana fermions with emergent SUSY as the low-energy effective model of an inhomogeneous 1D topological superconductor in the presence of periodically-modulated fields~\cite{klinovaja_transition_2012,kjaergaard_majorana_2012,ojanen_majorana_2013,maurer_designing_2018,marra_controlling_2017,marra_topologically_2019}.
This can be achieved in mesoscopic condensed matter systems, e.g., in semiconducting nanowire proximitized with a conventional superconductor.
By applying an external magnetic field, a proximitized semiconducting nanowire with Rashba spin-orbit coupling can be driven in a topologically nontrivial state, which exhibits Majorana modes localized at its ends.
In the presence of spatially-modulated magnetic fields, the chemical potentials, or superconducting pairing, the topological mass gap $\mathcal M$ can assume alternatively positive and negative values along the system, which correspond to topologically trivial and nontrivial segments.
Suppose the modulation is periodic with a periodicity $\lambda$ larger (but comparable) with the localization length of the Majorana end modes $\xi_\text{M}$.
In that case, one can realize a regime where Majorana end modes localize at the boundaries between trivial and nontrivial segments, realizing a 1D periodic lattice of partially overlapping Majorana modes.
In inhomogeneous 1D superconducting nanowires, Majorana modes localized at a finite distance and with finite overlaps~\cite{chevallier_mutation_2012,liu_zerobias_2012,kells_nearzeroenergy_2012,stanescu_disentangling_2013,roy_topologically_2013,san-jose_majorana_2016,liu_andreev_2017,liu_distinguishing_2018,moore_quantized_2018,ricco_spindependent_2018,chen_ubiquitous_2019,pan_generic_2019} are sometimes referred to in the literature as quasi-Majorana modes~\cite{prada_andreev_2020}.
The low-energy (infrared) effective theory of the system is described in terms of 0D Majorana modes on a 1D chain and with finite nearest-neighbor couplings, as in \cref{fig:chain}.
In the continuum, this corresponds to two counterpropagating 1D Majorana modes with finite dispersion, delocalized on the whole 1D system, and obeying an effective Dirac equation with a mass gap that can be controlled by externally applied magnetic or gate fields.
Notably, the two counterpropagating 1D Majorana modes appear as single-particle states below the bulk gap, i.e., separated from the higher-energy bulk states.
We identify these gapless Majorana modes as Goldstinos, i.e., the Nambu-Goldstone fermions~\cite{volkov_is_1973} associated with spontaneously broken SUSY\@.

This work is organized as follows.
In \cref{sec:chain}, we will briefly review the low-energy effective model (Majorana chain) describing Majorana modes localized on a 1D closed chain, and show how this model can describe dispersive massless or massive 1D Majorana modes.
In \cref{sec:susy}, we will show that these Majorana modes can be identified as Goldstinos.
Consequently, in \cref{sec:pwave}, we will show how the effective model introduced in the previous sections can be obtained as the low-energy limit of a 1D spinless topological superconductor with spatially-modulated fields.
In \cref{sec:swave}, we will describe a realistic implementation of the low-energy model in a quantum nanowire with Rashba spin-orbit coupling and proximitized with a conventional superconductor in the presence of spatially-modulated magnetic field or chemical potential.

\section{Majorana modes on a closed chain \label{sec:chain}}

\subsection{Majorana chain}

\begin{figure}[t]
\centering
\includegraphics[width=\columnwidth]{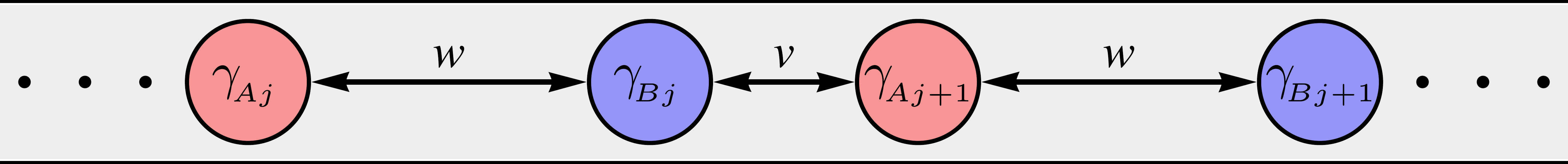}
\caption{
Majorana modes $\gamma_{Aj}$ and $\gamma_{Bj}$ localized on a 1D chain with nearest-neighbor couplings $v$ and $w$.
}
\label{fig:chain}
\end{figure}

\begin{figure}[tbp]
\centering
\includegraphics[width=\columnwidth]{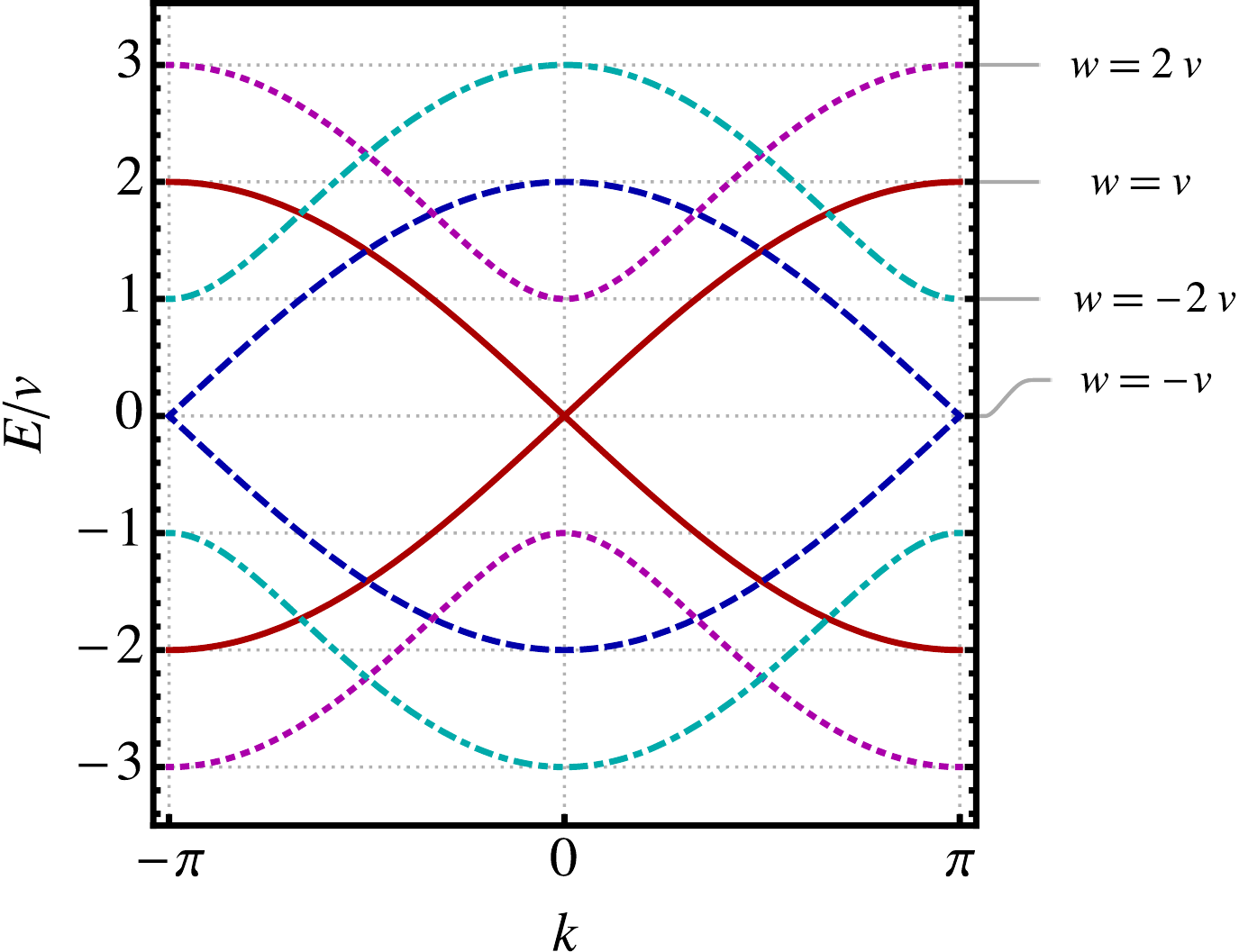} 
\caption{
Energy dispersion of a closed chain of Majorana modes $E_k=\sqrt{w^2 + v^2 - 2 w v \cos{k}}$ for different choices of $w/v$.
The dispersion is parabolic near the high-symmetry points $k=0,\pi$, except for the cases $w=v$, with linear dispersion at $k=0$ and $w=-v$, with linear dispersion at $k=\pi$.
}
\label{fig:chaindispersion}
\end{figure}

We consider an even number $2N$ of Majorana modes $\gamma_{Aj}$ and $\gamma_{Bj}$ on a closed chain as in \cref{fig:chain}, with each Majorana mode coupled to its nearest-neighbors, described by the Hamiltonian
\begin{equation}\label{eq:H}
\mathcal{H}_\text{eff}=
\ii \sum_{j=1}^N 
\left(
w \gamma_{Aj} \gamma_{Bj} + v \gamma_{Bj} \gamma_{Aj+1}
\right)
=\ii \Gamma A \Gamma^\intercal,
\end{equation}
with $w,v\in\mathbb R$ (i.e., we assume that the Hamiltonian is Hermitian), $\Gamma=(\gamma_{A1},\gamma_{B1},\dots,\gamma_{AN},\gamma_{BN})$, and where the Majorana operators satisfy $\gamma_{\alpha j}^\dag=\gamma_{\alpha j}$ and $\left\{\gamma_{\alpha i},\gamma_{\beta j}\right\}=2\delta_{ij}\delta_{\alpha\beta}$.
The system is translational invariant and forms a bipartite 1D lattice of $N$ unit cells, with two inequivalent Majorana modes per unit cell, respectively on the two sublattices $A$ and $B$.
This Hamiltonian can be generalized into the so-called Majorana-Hubbard model with the inclusion of interactions~\cite{rahmani_emergent_2015,rahmani_phase_2015,rahmani_interacting_2019}.
Since we consider a closed ring, we use the convention that the lattice sites $j=N+1$ and $j=1$ coincide.
This model can be regarded as the ``Majorana'' analogous of the Su-Schrieffer-Heeger model for polyacetylene~\cite{su_solitons_1979,su_soliton_1980}, when one substitutes the Dirac fermionic operators with Majorana operators $c_j\to\gamma_{\alpha j}$.
Moreover, it can be generalized to the translational-symmetry breaking Hamiltonian introduced in Ref.~\onlinecite{huang_supersymmetry_2017}, when one considers site-dependent hoppings $w\to w_j$ and $v\to v_j$.
Furthermore, the model can be reduced to a special case of the well-known Kitaev chain model describing a topological nontrivial superconducting state in 1D~\cite{kitaev_unpaired_2001}.
To show this, we define the Dirac fermionic operators $c_j=(\gamma_{Bj}+\ii \gamma_{Aj})/2$ and rewrite the Hamiltonian in Eq.~\eqref{eq:H} as
\begin{equation}\label{eq:Hfermion}
\mathcal{H}_\text{eff}=
-\sum_{j=1}^N 
\left[
2 w c_j^\dag c_j 
-
v 
\left(
c_j^\dag c_{j+1} 
+
c_j c_{j+1} + \text{h.c.}
\right)
\right],
\end{equation}
up to a constant term.
Thus the Majorana chain is equivalent to the Kitaev chain model~\cite{kitaev_unpaired_2001} if one takes $\mu =2w$, and $t=\Delta=-v$.
The effective Hamiltonian $\mathcal H_\text{eff}$ is in the BDI symmetry class~\cite{kitaev_periodic_2009,schnyder_classification_2009,ryu_topological_2010,ludwig_topological_2016} with unbroken particle-hole, time reversal, and chiral symmetries.
Many properties of Hamiltonian in Eq.~\eqref{eq:H} can be directly derived from the properties of the well-studied Kitaev chain model~\cite{kitaev_unpaired_2001}.
In order to be self-contained and set the stage for the discussion of SUSY, we will briefly review these properties and translate them into the language of the Majorana model in Eq.~\eqref{eq:H}.
By Fourier transform, $c_j=(1/\sqrt{N})\sum_k \ee^{\ii k j} c_k$, the Hamiltonian in momentum space becomes
\begin{equation}
\thinmuskip=.5\thinmuskip\medmuskip=.5\medmuskip\thickmuskip=.5\thickmuskip
\mathcal{H}_\text{eff}=
\sum_k 
\left[
2\left( v\cos{k} - w\right) c^\dag_k c_k 
+
\left( \ii v \sin{k}\ c^\dag_{k} c^\dag_{-k} +\text{h.c.} \right)
\right],
\end{equation}
where the momenta are quantized as $k=2\pi m/N$ with $m=1,\ldots,N$ integer.
The model is symmetric upon changing the signs of the parameters $v$ and $w$, for example, $v\to-v$ by redefining the momentum as $k\to k+\pi$, or equivalently by using the unitary transformation $c_j\to (-1)^j c_j$ in real space.
We can rewrite the Hamiltonian in Bogoliubov-de~Gennes form as
\begin{align}\label{eq:Hk}
\mathcal{H}_\text{eff}&=
\sum_k
[c_k^\dag , c_{-k}]
\cdot
H(k)
\cdot
\begin{bmatrix} c_k \\ c_{-k}^\dag\\ \end{bmatrix},
\nonumber\\
\text{with \, }
H(k)&= (v\cos{k} -w)\tau_z + v\sin{k}\,\tau_y,
\end{align}
up to a constant term, and where $\boldsymbol\tau=(\tau_x,\tau_y,\tau_z)$ are the Pauli matrices in particle-hole space.
The Hamiltonian in momentum space can be written in the form $H(k)=\mathbf{H}(k)\cdot\boldsymbol\tau$, where $\mathbf{H}(k)$ is the Anderson pseudovector.
The Hamiltonian is diagonalized as
\begin{equation}\label{eq:HBdGk}
\mathcal{H}_\text{eff}=
\setlength{\arraycolsep}{1pt}
\sum_k
[d_k^\dag , d_{-k}]
\cdot
\begin{bmatrix}
E_k & 0\\
0 & -E_k\\
\end{bmatrix}
\cdot
\begin{bmatrix} d_k \\ d_{-k}^\dag\\ \end{bmatrix}
= \sum_k 2E_k d_k^\dag d_k,
\end{equation}
up to a constant term, where $d_k$ are a new set of fermionic operators with dispersion
\begin{equation}\label{eq:chaindispersion}
E_k=\sqrt{w^2 + v^2 - 2 w v \cos{k}},
\end{equation}
where positive and negative energy levels correspond respectively to particle and hole states.
The eigenstates operators $d_k=\alpha_k c_k + \beta_k c_{-k}^\dag$ are defined by the unitary Bogoliubov transformation
\begin{equation}\label{eq:Bogoliubov1}
\begin{bmatrix}
d_k \\
d_{-k}^\dag\\
\end{bmatrix}=
U_k
\cdot
\begin{bmatrix}
c_k \\
 c_{-k}^\dag\\
\end{bmatrix},
\qquad 
\text{with \, }
U_k=
\begin{bmatrix}
\alpha_k & \beta_k \\
-\beta_k^* & \alpha_k^* \\
\end{bmatrix},
\end{equation}
where the coefficients $\alpha_k$ and $\beta_k$ for $E_k\neq0$ are given by
\begin{subequations}\label{eq:alphabeta}\begin{align}
\alpha_k=&
\sqrt{\frac12 + \frac{v\cos{k} - w}{2 E_k}},
\\\quad
\beta_k=-\ii \,\sgn{\left(v \sin{k}\right)} 
&
\sqrt{\frac12 - \frac{v\cos{k} - w}{2 E_k}},
\end{align}\end{subequations}
up to an overall phase.
The energy dispersion is shown in \cref{fig:chaindispersion} for different values of the parameters $v,w$.
The dispersion has a mass gap $\mathcal M_\text{eff}= |w|-|v|$, being either parabolic (if $|v|\neq|w|$) or linear (if $|v|=|w|$) near the high-symmetry point $k=0$ (if $v w>0$) or $k=\pi$ (if $v w<0$).
For $|v|\neq|w|$, the dispersion is always gapped $E_k\neq0$.
In particular, the minimum and maximum energies are reached at the time-reversal symmetry points $k=0,\pi$, being $E_\text{min}=\min (E_0,E_\pi)=|\mathcal M_\text{eff}|$ and $E_\text{max}=\max (E_0,E_\pi)$, with $E_0=|v-w|$ and $E_\pi=|v+w|$.

\subsection{1D Majorana modes and 0D end modes}

In the massless case $|v|=|w|$, the dispersion in Eq.~\eqref{eq:chaindispersion} becomes
\begin{align}
E_k&=
\begin{cases}
2\left\vert v \sin{\frac k2}\right\vert & \text{\, for \, } v=w, \\[.5em]
2\left\vert v \cos{\frac k2}\right\vert & \text{\, for \, } v=-w. \\
\end{cases}
\end{align}
The gap closes linearly at one of the two symmetry points $k=0,\pi$, with $E_0=0$ or $E_\pi=0$ for $v=w$ and $v=-w$, respectively.
Notice that if the number of unit cells $N$ is even, the quantized momenta $k=2\pi m/N$ include both symmetry points $k=0$ and $\pi$, whereas if $N$ is odd, they include only $k=0$.
Therefore, for $N$ even, the gap closes for $v=\pm w$, whereas for $N$ odd, the gap closes only for $v=w$.
The Hamiltonian becomes $H(0)=0$ and $H(\pi)=0$ respectively for $v=\pm w$.
Thus the zero-energy eigenstates at the gapless points are doubly degenerate and given by any combination of the fermionic operator $c_{k=0}=(1/\sqrt N) \sum_n c_n$ and its Hermitian conjugate $c_{k=0}^\dag$ for $v=w$, or by any combination of $c_{k=\pi}=(1/\sqrt N) \sum_n (-1)^n c_n$ and $c_{k=\pi}^\dag$ for $v=-w$.
Alternatively, the eigenstates can be written as a superposition of a fermionic operator $d_\text{M}$ and its hermitian conjugate $d_\text{M}^\dag$, defined as
\begin{equation}\label{eq:nonlocalfermion}
d_\text{M}=\frac12\left( {\widetilde\gamma}_A+\ii{\widetilde\gamma}_B \right),
\end{equation}
where the nonlocal Majorana operators ${\widetilde\gamma}_{A,B}$ are defined as
\begin{equation}\label{eq:nonlocalmajo}
{\widetilde\gamma}_A=\frac1{\sqrt{N}}
\sum_{j=1}^N \ee^{\ii k j} \gamma_{Aj},
\qquad
{\widetilde\gamma}_B=\frac1{\sqrt{N}}
\sum_{j=1}^N \ee^{\ii k j} \gamma_{Bj},
\end{equation}
which satisfy
${\widetilde\gamma}_{\alpha}^\dag={\widetilde\gamma}_{\alpha}$ and
$\left\{{\widetilde\gamma}_{\alpha},{\widetilde\gamma}_{\beta}\right\}=2\delta_{\alpha\beta}$,
and with $k=0,\pi$ respectively for $v=w$ and $v=-w$ [also notice that $\ee^{\ii k j}=(w/v)^j$].
These are two delocalized 1D Majorana modes with finite support on the whole chain.
The model exhibits two topologically phases corresponding to the topological invariant ${\mathcal P}_\text{eff}=\sgn\pf A$ where $\pf A=w^2-v^2$, which gives ${\mathcal P}_\text{eff}=\sgn{(|w|-|v|)}=\sgn{\mathcal M_\text{eff}}$.
In the massless case, the 1D Majorana modes ${\widetilde\gamma}_{A,B}$ describe a 1D free Majorana fermion which is equivalent to the 1+1D Ising model (via the Jordan-Wigner transformation) and thus described by a 1+1D conformal field theory in the Ising universality class with the central charge $c=1/2$~\cite{sato_majorana_2016}.

Besides 1D Majorana modes, the Hamiltonian also exhibits Majorana end modes in the nontrivial phase $|v|>|w|$ in the case of open boundary conditions.
Up to exponentially small corrections, the two 0D end states are
\begin{subequations}\label{eq:endstates}\begin{align}
{\widetilde\gamma}_\text{L}&=
\sqrt{\frac{v^2-w^2}{w^2}}\sum_{j=1}^N \left(\frac wv\right)^j \gamma_{Aj},
\\
{\widetilde\gamma}_\text{R}&=
\sqrt{\frac{v^2-w^2}{w^2}} \sum_{j=1}^N \left(\frac wv\right)^{N+1-j} \gamma_{Bj},
\end{align}\end{subequations}
localized respectively on the left and right ends of the chain.
The two 0D end states have a localization length $\xi_\text{eff}=1/|\ln|w/v||$ and form a nonlocal fermionic state $d_\text{M}=({\widetilde\gamma}_\text{L}+\ii{\widetilde\gamma}_\text{R})/2$ with exponentially small energy $\propto \ee^{-N/\xi_\text{eff}}$.
Notice that the energy of Majorana end modes is always finite (although exponentially small) for $w\neq0$, since in the presence of open boundary conditions one can verify that $\det A=w^N\neq0$.

In the presence of disorder, the translational symmetry of the Hamiltonian in Eq.~\eqref{eq:H} breaks down, and the model can be generalized by considering site-dependent hoppings $v,w\to v_j,w_j$, as in Ref.~\onlinecite{huang_supersymmetry_2017}.
In this case, the particle-hole gap closes if and only if $\det{A}=\big(\prod_j v_j - \prod_j w_j\big)^2=0$, i.e., if and only if
\begin{equation}\label{eq:gaplesscondition}
\prod_{j=1}^N v_j=\prod_{j=1}^N w_j,
\end{equation} 
as derived in Ref.~\onlinecite{huang_supersymmetry_2017}.
Without loss of generality, one can take $v_j= v +\delta v_j$ and $w_j= w +\delta w_j$, where $v$ and $ w$ are here the average couplings while $\delta v_j$ and $\delta w_j$ the small random corrections due to disorder.
Hence, the nonlocal zero-energy fermionic state can be written again as in Eq.~\eqref{eq:nonlocalfermion}, but with the nonlocal Majorana operators now defined as
\begin{equation}\label{eq:nonlocalmajodisorder}
{\widetilde\gamma}_A=
\frac1{|\phi_A|}
\sum_{j=1}^N
\phi_{Aj}
\gamma_{Aj},
\quad
{\widetilde\gamma}_B=
\frac1{|\phi_B|}
\sum_{j=1}^N
\phi_{Bj}
\gamma_{Bj},
\end{equation}
with $\phi_{A}=(1,w_1/v_1,w_1w_2/v_1v_2,\dots,\prod_{j=1}^{N-1}w_j/v_j)$ and $\phi_{B}=(1,v_1/w_1,v_1v_2/w_1w_2,\dots,\prod_{j=1}^{N-1}v_j/w_j)$, as derived in Ref.~\onlinecite{huang_supersymmetry_2017}.
The topological invariant can be still defined in the presence of disorder, as the sign of the Pfaffian~\cite{kitaev_unpaired_2001} of the Hamiltonian in real space expressed in the Majorana basis as in Eq.~\eqref{eq:H}, that is
\begin{equation}
{\mathcal P}_\text{eff}=\sgn\pf A=
\sgn\left(
\prod_{j=1}^N w_j
-
\prod_{j=1}^N v_j
\right).
\end{equation}
The closing of the particle-hole gap occurs with a change of the topological invariant when $\pf A=0$, i.e., when $\prod_j w_j - \prod_j v_j=0$.
The Majorana end modes localized at the two ends of the chain in the topologically nontrivial phase are robust against disorder.

In the presence of disorder, SUSY is realized if and only if Eq.~\eqref{eq:gaplesscondition} is satisfied.
Even in this case, SUSY is realized at least one point in the parameter space, which approximately corresponds to $|v|\approx|w|$ in the case of small disorder.
Indeed, if $|v|\gg|w|$, one has ${\mathcal P}_\text{eff}=-1$, while if $|v|\ll|w|$, one has ${\mathcal P}_\text{eff}=1$.
Hence, there exists at least one value $v^*$ (or equivalently $w^*$) where the $\prod_j v_j - \prod_j w_j=0$ and Eq.~\eqref{eq:gaplesscondition} is satisfied.
Consequently, SUSY occurs at least at a single point of the parameter space.
Since we consider small disorder (i.e., $|v|\gg|\delta v_j|$ and $|w|\gg|\delta w_j|$), this solution must be $ v^*\approx\pm w$.
An important consequence is that changing the disorder configuration (i.e., changing the values of $\delta v_j$ and $\delta w_j$) does not remove the SUSY point from the parameter space, but merely changes the value of $ v^*$ at which SUSY occurs.

\subsection{Equivalence between 1D Majorana modes and Majorana fermions}

The continuum limit of Hamiltonian in Eq.~\eqref{eq:Hk} can be directly obtained in momentum space by using the approximations $\sin{k}\approx k$, and $\cos{k}\approx 1-{k^2}/2$ (assuming $vw>0$), which give
\begin{equation}
H= v k \,\tau_y + \left(m v^2-\frac{v}2 k^2\right)\tau_z,
\label{eq:HkDirac}
\end{equation}
which coincides with a 1D Dirac equation with mass $m =(v-w)/v^2$ (i.e., mass gap $m v^2=v-w$) and a quadratic correction in the momentum.
The energy dispersion is
\begin{equation}
E_k=\sqrt{\left(mv^2-\frac{v}2 k^2\right)^2 + v^2 k^2},
\end{equation}
which for $w=v$ becomes linear at small momenta $E_k\approx v|k| $.
The solutions of the Dirac equation above are described by two counterpropagating and degenerate eigenfunctions, i.e., with opposite momenta and same energy $E_k=E_{-k}$.
The covariant form of the Dirac equation is obtained by multiplying Eq.~\eqref{eq:HkDirac} by $\tau_z$, which in real space reads
\begin{equation}\label{eq:DiracCovariant}
H= v \tau_x \partial_x + \left(m v^2-\frac{v}2 p^2\right),
\end{equation}
which is a real equation with real solutions.
Dirac equations with a quadratic correction in the form of Eq.~\eqref{eq:DiracCovariant} are topologically trivial for $mv<0$ or nontrivial for $mv>0$~\cite{shen_topological_2011}.
Consequently, the Dirac equation is trivial or nontrivial respectively for $\mathcal M_\text{eff}>0$ and $\mathcal M_\text{eff}<0$, where $\mathcal M_\text{eff}=|w|-|v|$ is the topological mass gap, assuming $v w>0$.
The case $vw<0$ can be obtained using the transformation $v\to-v$ by redefining the momentum as $k\to k+\pi$.

\subsection{Pseudohelicity and Majorana pseudospin}

We define the Majorana pseudospin operator as $\boldsymbol\tau/2$ in analogy to the spin operator $\boldsymbol\sigma/2$ (we use natural units).
Here, $\boldsymbol\sigma$ and $\boldsymbol\tau$ are the vectors of Pauli matrices in spin and particle-hole space, respectively.
Majorana fermions in particle physics are obtained by decoupling the Dirac four-component bispinor into two independent right-handed and left-handed two-component spinors (chiral projections).
The expectation values of the two operators calculated for each chiral projection $\Psi_\text{L,R}$ coincide, i.e., $\langle \boldsymbol \sigma \rangle_\text{L,R} =\langle \boldsymbol \tau \rangle_\text{L,R}$.
In the massless case, each chiral projection forms a pair of counterpropagating modes, which are symmetric under particle-hole and time-reversal symmetry.
As a consequence of time-reversal symmetry, these modes have opposite spin, i.e., they form a helical pair of counterpropagating chiral modes.
Moreover, due to particle-hole symmetry, they have opposite Majorana pseudospin (opposite particle-hole content), i.e., they form a \emph{pseudohelical} pair.
Analogously to the notion of helical modes, i.e., a pair of counterpropagating modes having opposite spin, we define pseudohelical modes as pair of counterpropagating modes having opposite pseudospin.
Since the expectation values of spin and Majorana pseudospin coincide in the case of spinful and massless Majorana fermions, the two counterpropagating modes are both helical and pseudohelical.
However, in the case of spinless particles, pseudochirality and pseudohelicity are still well-defined, whereas chirality and helicity are not.

The eigenstates of the effective Hamiltonian $\mathcal H_\text{eff}$ in Eq.~\eqref{eq:alphabeta} have expectation values of the Majorana pseudospin given by
\begin{align}
\langle \boldsymbol\tau/2\rangle &=
\left(
\Re{( \alpha_k \beta_k^*}),\
\Im{( \alpha_k \beta_k^*}),\
\frac12\left(|\alpha_k|^2- |\beta_k|^2\right)
\right)=
\nonumber\\&=
\left(0,\,
\frac{v\sin{k}}{2E_k}, \
\frac{v\cos{k} -w}{2E_k}
\right)
=\frac{\mathbf{H}(k)}{2E_k}.
\end{align}
In the massless case $v=w$ and at $k\approx0$ (or alternatively for $v=-w$ and at $k\approx\pi$), one obtains $\langle \boldsymbol\tau/2 \rangle =\sgn{(v\sin{k})}(0,1/2,0)$.
In particular, for $v=w>0$ one has $\langle \boldsymbol\tau/2 \rangle =\pm(0,1/2,0)$ for $k\to 0^\pm$, while for $v=-w>0$ one has $\langle \boldsymbol\tau/2 \rangle =\pm(0,1/2,0)$ for $k\to \pm\pi$.
Hence, in the massless case, the effective Hamiltonian $\mathcal H_\text{eff}$ exhibits a pair of counterpropagating modes with opposite Majorana pseudospin.
The projections of the Majorana pseudospin on the momentum direction are the same for both modes, i.e., the two modes form a pseudohelical pair with a well-defined pseudochirality.
Consequently, elastic backscattering between the two branches is suppressed.
The Majorana pseudospin $\boldsymbol\tau/2$ introduced here is a generalization of the Majorana polarization~\cite{sticlet_spin_2012,bena_testing_2017,marra_majorana/andreev_2021}:
The expectation values of the first two components of the Majorana pseudospin coincide with the Majorana polarization (up to a $1/2$ prefactor).
Thus the pseudohelical pair of counterpropagating modes have opposite Majorana pseudospin and opposite Majorana polarization.
The notion of pseudohelicity introduced here must not be confused with the related but different notion of pseudohelical edge states in graphene~\cite{tobias_protected_2018}.

\section{Emergent quantum mechanical SUSY \label{sec:susy}}

A 1D closed chain of 0D Majorana modes exhibits SUSY, as shown in the case of translationally invariant systems~\cite{hsieh_all_2016} and generic closed chains without translational invariance~\cite{huang_supersymmetry_2017}.
Hereafter, we will shortly recall these known results and specialize them to our specific model.
The groundstate of the Bogoliubov-de~Gennes Hamiltonian in Eq.~\eqref{eq:HBdGk} is the fermion vacuum~\cite{read_paired_2000}, i.e., the state $\ket{0}$ which annihilated by all fermionic quasiparticle operators $d_k$, that is, $d_k\ket{0}=0$, $\forall k$, which can be written explicitly as $\ket{0}=\prod_{0<k<\pi} d_{-k}d_k\ket{\varnothing}$, where $\ket{\varnothing}$ is the vacuum of the bare fermionic operators (i.e., $c_k\ket{\varnothing}=0$, $\forall k$).
The many-body spectra of the Hamiltonian are given by the quasiparticle excitations on top of the fermionic condensate.
When the 1D Majorana fermion is massless $|v|=|w|$, the spectra have the remarkable property that all many-body energy levels are at least doubly degenerate.
This is the direct consequence of the presence of the zero-energy Dirac fermion $d_\text{M}$ defined in Eq.~\eqref{eq:nonlocalfermion}.
Indeed, for any many-body state $\ket{\psi}$, there exists at least another state with the same energy $\ket{\psi'}$, distinguished by a different value of the zero-energy single-particle occupation number $n_\text{M}=0,1$.
Specifically, 
$\ket{\psi'}=d_\text{M}^\dag\ket{\psi}$ if $n_\text{M}\ket{\psi}=0$
and
$\ket{\psi'}=d_\text{M}\ket{\psi}$ if $n_\text{M}\ket{\psi}=\ket{\psi}$.
Alternatively, by using the Majorana operators, for any many-body state $\ket{\psi}$ there exists at least another state ${\widetilde\gamma}_{A,B}\ket{\psi}$ with the same energy.
Moreover, the two degenerate states are distinguished by a different fermion parity.
This is because creating or annihilating the zero-energy quasiparticle fermion $d_\text{M}$ does not change the total energy but changes the total number of fermions by one, and therefore changes the fermion parity.
We define the fermion parity operator $P=(-1)^{F}=\prod_{j=1}^{N}\ii\gamma_{Aj}\gamma_{Bj}$, where $F$ is the fermion number operator.
The parity operator commutes with the Hamiltonian and anticommutes with the Majorana operators $\{{\widetilde\gamma}_{A,B},P\}=0$.
Hence, the many-body states $\ket{\psi}$ and ${\widetilde\gamma}_{A,B}\ket{\psi}$ have opposite fermion parity.

The twofold degeneracy of the many-body energy spectra can be explained as the consequence of SUSY in quantum mechanics.
In a nutshell, SUSY can be defined as a symmetry with respect to the change of the fermion parity of the many-body state~\cite{qi_time-reversal-invariant_2009,witten_dynamical_1981,cooper_supersymmetry_1995,gangopadhyaya_supersymmetric_2017}.
To make all energy levels nonnegative, we shift the Hamiltonian by a positive constant and define
\begin{equation}
{\mathcal H}_\text{SUSY}=
{\mathcal H}_\text{eff} + h(|v|+|w|)
\end{equation}
with $h>1$.
We will show that the Hamiltonian ${\mathcal H}_\text{SUSY}$ exhibits centrally extended quantum mechanical SUSY\@.
In general, a many-body Hamiltonian ${\mathcal H}$ exhibits ${\mathcal N}=2$ SUSY~\cite{witten_dynamical_1981,cooper_supersymmetry_1995,gangopadhyaya_supersymmetric_2017} with zero superpotential if there exist two supercharges $Q_1,Q_2$ which commute with the Hamiltonian and satisfy the superalgebra
\begin{equation}\label{eq:hermitiansuperalgebra}
\{ P, Q_i \}=0,\qquad \{ Q_i, Q_j \}=2 \delta_{ij} {\mathcal H}.
\end{equation}
One can also combine the supercharges $Q_1,Q_2$ into a complex supercharge $Q$ which commutes with the Hamiltonian and satisfy the superalgebra
\begin{equation}\label{eq:nonhermitiansuperalgebra}
\{ P, Q \}=0, \quad
Q^2=( Q^\dag)^2=0, \quad
\{ Q, Q^\dag\}=2{\mathcal H},
\end{equation}
where $Q=(Q_{1}+\ii Q_{2})/\sqrt2$.
The definitions of SUSY via the superalgebras in Eqs.~\eqref{eq:hermitiansuperalgebra}~and~\eqref{eq:nonhermitiansuperalgebra} are equivalent.

\subsection{SUSY and gapless Majorana mode}

Following Ref.~\onlinecite{huang_supersymmetry_2017}, we can define two hermitian operators
\begin{equation}\label{eq:supercharges}
Q_{A,B}=\sqrt{{\mathcal H}_\text{SUSY}}\,{\widetilde\gamma}_{A,B},
\end{equation}
where ${\widetilde\gamma}_{A,B}$ are the nonlocal Majorana operators defined in Eq.~\eqref{eq:nonlocalmajo}.
The hermitian operators $Q_{A,B}$ commute with the Hamiltonian and satisfy the superalgebra in Eq.~\eqref{eq:hermitiansuperalgebra}.
The two supercharges can be combined into a complex supercharge
\begin{align}\label{eq:Msupercharge}
Q_\text{M}=
\sqrt{\frac{{\mathcal H}_\text{SUSY}}2}\,( {\widetilde\gamma}_A+\ii {\widetilde\gamma}_B)
=\sqrt{2{\mathcal H}_\text{SUSY}}\,d_\text{M},
\end{align}
which satisfies the superalgebra in Eq.~\eqref{eq:nonhermitiansuperalgebra}.
Hence, the Hamiltonian exhibits at least ${\mathcal N}=2$ SUSY with supercharges $Q_\text{M},Q_\text{M}^\dag$.
The action of these supercharges on a generic many-body state is to annihilate or create the fermionic zero-energy state $d_\text{M}$ in Eq.~\eqref{eq:nonlocalfermion}.

SUSY explains the twofold degeneracy of the spectra.
For any many-body particle state $\ket{\psi}_\pm$ with definite fermion parity $P\ket{\psi}_\pm=\pm\ket{\psi}_\pm$, there exists a supersymmetric degenerate state (superpartner) defined as
\begin{equation}
\ket{\psi}_\mp=
\frac1{\sqrt{E_\psi}} \, Q_{A,B}\ket{\psi}_\pm,
\end{equation}
with opposite fermion parity.
The above expression is well-defined for all many-body particle states, since $E_\phi>0$.
In our case, the Hamiltonian ${\mathcal H}_\text{eff}$ is diagonal in the many-particle basis $\ket{\{n_m\}}=\ket{n_1,n_2,\dots,n_N}$, where $n_m=0,1$ are the occupation numbers of the fermionic single-particle state with momentum $k=2\pi m/N$.
Therefore the operator $\sqrt{{\mathcal H}_\text{SUSY}}$ is obtained by taking the square roots of the diagonal elements of ${\mathcal H}_\text{SUSY}$.
SUSY persists in the presence of disorder as long as the particle-hole gap is closed, i.e., if Eq.~\eqref{eq:gaplesscondition} is satisfied.
In this case, the supercharges in Eq.~\eqref{eq:supercharges} need to be redefined in terms of the nonlocal Majorana operators in Eq.~\eqref{eq:nonlocalmajodisorder}.
The emergence of SUSY with supercharges $Q_{A,B}$ relies only on the presence of a bulk zero-energy state closing the particle-hole gap.
Therefore this argument can be extended to any topological superconductor at the transition between trivial and nontrivial phase, e.g., a Kitaev chain with $|\mu|=2t$~\cite{kitaev_unpaired_2001}.

\subsection{SUSY and translational symmetry}

In the translational invariant case realized in the absence of disorder and with $|v|=|w|$, one can also consider a second SUSY superalgebra by defining a set of supercharges in terms of a translational symmetry operator, as derived in Ref.~\onlinecite{hsieh_all_2016}.
Assuming $v=w$, the Hamiltonian is invariant with respect to the translation operator $T$ defined by 
\begin{equation}
T \gamma_{Aj} T^\dag= \gamma_{Bj},\qquad
T \gamma_{Bj} T^\dag= \gamma_{Aj+1(\mathrm{mod} N)}.
\end{equation}
(The case $v=-w$ can be analyzed by using the unitary transformation $c_j\to (-1)^j c_j$, which gives $v\to-v$.)
The translation operator $T$ anticommutes with the fermion parity operator $\{T,P\}=0$, and commutes with the Hamiltonian $[T,{\mathcal H}_\text{SUSY}]=0$ in the case $|v|=|w|$ and in the absence of disorder.
We notice that the Hamiltonian in Eq.~\eqref{eq:H} is invariant up to the translation operator $T^2$ for any choice of $v,w$.
Thus one can define a complex supercharge
\begin{equation}\label{eq:Tsupercharge}
{Q_\text{T}}=\sqrt{\frac{{\mathcal H}_\text{SUSY}}2}\, T (1+P),
\end{equation}
which satisfies the superalgebra in Eq.~\eqref{eq:nonhermitiansuperalgebra}.
Given a many-body state $\ket{\psi}_\pm$ with parity $\pm1$, the degenerate superpartner with opposite parity is obtained respectively as $(1/\sqrt{2E_\psi}) Q_\text{T}\ket{\psi}_+$ and $(1/\sqrt{2E_\psi}) Q_\text{T}^\dag\ket{\psi}_-$.
Hence, the Hamiltonian exhibits ${\mathcal N}=2$ SUSY with supercharges $Q_\text{T},Q_\text{T}^\dag$.
It is also possible to decompose the complex supercharge $Q_\text{T}$ into two Hermitian supercharges which satisfy the superalgebra in Eq.~\eqref{eq:hermitiansuperalgebra}.
Moreover, the complex supercharge $Q_\text{T}$ can be written in terms of the supercharges $Q_{A,B}$ as
\begin{equation}\label{eq:TMrelation}
Q_\text{T}
= \frac1{\sqrt2} U_{A} Q_{A} (1+P)
= \frac1{\sqrt2} U_{B} Q_{B} (1+P),
\end{equation}
where $U_{A,B}=T\,\widetilde\gamma_{A,B}$ are unitary operators.

\subsection{Extended $\mathcal N=4$ SUSY with central charges}

In general, a supersymmetric Hamiltonian can admit several supercharges with a richer structure, which is known as \emph{extended} SUSY~\cite{niederle_extended_2001}.
An extended $\mathcal N = 2n$ superalgebra is a set of $2n$ supercharges $Q_i,Q_i^\dag$ which satisfy the following superalgebra
\begin{subequations}\label{eq:extendedsusy}
\begin{align}
\{ Q_i,Q_j^\dag \}&=2\delta_{ij}\mathcal{H}+{Z}_{ij},\qquad i,j=1,\ldots,n,
\label{eq:extendedsusya}
\\
\{ Q_i,Q_j \}&=\{ Q_i^\dag,Q_j^\dag \}=0,
\label{eq:extendedsusyb}
\end{align}
\end{subequations}
where ${Z}_{ij}$ are the so-called central charges, which commute with all elements $Q_i$, $Q_i^\dag$, and $\mathcal{H}$ of the superalgebra~\cite{niederle_extended_2001}.
It is natural to ask whether one can combine the supersymmetries described above into an extended SUSY\@.
The two complex supercharges
\begin{subequations}
\begin{align}\label{eq:supercharge1}
{\mathcal Q}_1
&
=\sqrt{\frac{{\mathcal H}_\text{SUSY}}2}\, d_\text{M} (1+P)
=\frac12 Q_\text{M} (1+P),
\\
\label{eq:supercharge2}
{\mathcal Q}_2
&
=\sqrt{\frac{{\mathcal H}_\text{SUSY}}2}\, T (1+P)
=Q_\text{T},
\end{align}
\end{subequations}
satisfy the centrally extended superalgebra in Eq.~\eqref{eq:extendedsusy} with central charges
\begin{equation}\label{eq:Zij}
{\mathcal Z}_{ij}=
{\mathcal H}_\text{SUSY}
\begin{bmatrix}
-(1+P') & \{ d_\text{M} (1+P), T^\dag \} \\
\{ d_\text{M}^\dag (1-P),T \} & 0\\
\end{bmatrix},
\end{equation}
where $P'=(-1)^{F-n_\text{M}}=P(1-2n_\text{M})=-\ii P\, \widetilde\gamma_A\widetilde\gamma_B$.
Hence, the Hamiltonian exhibits $\mathcal N = 4$ centrally extended SUSY\@.

\subsection{Spontaneous partial breaking of SUSY}

An important question is whether the SUSY is spontaneously broken.
To see this, we consider the Witten index~\cite{witten_dynamical_1981,witten_constraints_1982} defined as ${\mathcal W}=\Tr(-1)^{F}$, where $F$ is the fermion number operator and the trace is extended to all many-body eigenstates of $\mathcal{H}_\text{SUSY}$.
The Witten index coincides with the difference between the number of eigenstates with even and odd fermion parity, and it is ${\mathcal W} = 0$ for spontaneously broken SUSY and ${\mathcal W} = 1$ for unbroken SUSY\@.
The groundstate is doubly degenerate, and all many-body eigenstates, including the groundstate, have a supersymmetric partner with opposite parity (being at least double-degenerate).
Therefore the Witten index is zero, which corresponds to spontaneously broken SUSY and mandates the presence of a Goldstino, i.e., a Nambu-Goldstone fermion~\cite{volkov_is_1973}.

The spontaneously breaking of SUSY can be interpreted as the breaking of the extended ${\mathcal N}=4$ SUSY down to ${\mathcal N}=2$ SUSY\@.
To show this, let's consider the action of the supercharges on the two degenerate groundstates, i.e., the vacuum $\ket{0}$ and the state $\ket{1}=d_\text{M}^\dag\ket{0}$, having respectively even and odd fermion parity.
The supercharges ${\mathcal Q}_1,{\mathcal Q}_1^\dag$ annihilate both groundstates
\begin{equation}
{\mathcal Q}_1\ket{0} = {\mathcal Q}_1\ket{1} = 
{\mathcal Q}_1^\dag\ket{0}=
{\mathcal Q}_1^\dag\ket{1}=0.
\end{equation}
However, the two groundstates are superpartners with respect to the supercharges ${\mathcal Q}_2,{\mathcal Q}_2^\dag$.
Indeed, since $T$ commutes with the Hamiltonian and anticommutes with the parity operator, the eigenstate $T\ket{0}$ has the same energy but opposite parity of $\ket{0}$, and therefore $T\ket{0}=\ket{1}$ and also $T\ket{1}=\ket{0}$, which gives
\begin{subequations}\label{eq:partiallybroken}
\begin{align}
{\mathcal Q}_2\ket{0} = 
&
\sqrt{2{\mathcal H}_\text{SUSY}}\, \ket{1},
\qquad
{\mathcal Q}_2^\dag\ket{1} = 
\sqrt{2{\mathcal H}_\text{SUSY}}\, \ket{0},
\\
{\mathcal Q}_2^\dag\ket{0}=
&
{\mathcal Q}_2\ket{1}=0.
\end{align}
\end{subequations} 
Hence, the supercharge ${\mathcal Q}_1$ remains unbroken while the supercharge ${\mathcal Q}_2$ is spontaneously broken resulting in a Goldstino (or Nambu-Goldstone fermion)~\cite{volkov_is_1973}.
More precisely, the Hamiltonian possesses ${\mathcal N}=4$ SUSY $({\mathcal Q}_1 , {\mathcal Q}_2)$ spontaneously broken down to ${\mathcal N}=2$ SUSY ${\mathcal Q}_1$.
Thus, we identify the gapless Majorana fermion as the Goldstino of the spontaneously broken SUSY\@.

In extended superalgebras with zero central charges ${\mathcal Z}_{ij}=0$, the partial supersymmetry breaking is forbidden by the no-go theorem by Witten~\cite{witten_dynamical_1981,witten_constraints_1982}.
This means that, in this case, either all supersymmetries $\mathcal{Q}_i$ are broken or they are all unbroken.
However, the no-go theorem can be evaded in the presence of nonzero central charges~\cite{ivanov_partial_1991}, as in Eq.~\eqref{eq:Zij}.
This leads to the partial breaking of the extended SUSY in Eq.~\eqref{eq:partiallybroken}.

There have been several proposals for realizing SUSY in condensed matter systems, such as 
in Bose-Fermi mixtures of ultracold atoms~\cite{snoek_ultracold_2005,snoek_theory_2006,yu_supersymmetry_2008,yu_simulating_2010,shi_supersymmetric_2010,lai_relaxation_2015,blaizot_spectral_2015,bradlyn_supersymmetric_2016,blaizot_goldstino_2017,tajima_goldstino_2021}, in discrete Majorana chains~\cite{rahmani_emergent_2015,rahmani_phase_2015,rahmani_interacting_2019}, Majorana Cooper-pair boxes~\cite{ebisu_supersymmetry_2019}, and at the boundary of topological superconductors and insulators~\cite{grover_emergent_2014,ponte_emergence_2014,ma_realization_2021}.
However, our proposal gives the first condensed matter realization of an extended SUSY with central extensions, which plays essential roles in high energy theory to study non-perturbative aspects of quantum field theory~\cite{Witten:1978mh,Seiberg:1994rs,Seiberg:1994aj}.

\subsection{Effect of disorder and open boundary conditions}

In the absence of disorder, the translational invariance is a necessary and sufficient condition for closing the mass gap $\mathcal M_\text{eff}=0$, and it is also a necessary and sufficient condition for the emergence of SUSY\@.
The mass gap is zero if and only if the Hamiltonian exhibits SUSY:
In other words, the closing of the mass gap is protected by SUSY\@.
For systems with disorder and no translational symmetry, a necessary and sufficient condition for the closing of the mass gap and the emergence of SUSY is given by Eq.~\eqref{eq:gaplesscondition}.
Hence, also in this case, the closing of the mass gap is protected by SUSY\@.

Finally, we consider the case of a 1D chain with open boundary conditions.
In this case, the Hamiltonian still exhibits SUSY in the limit of infinite system size, or in the case where one of the overlaps is zero $w=0$ (or if any $w_{n}=0$, such that Eq.~\eqref{eq:gaplesscondition} is satisfied, in the presence of disorder).
In these two cases, the Majorana end modes ${\widetilde\gamma}_\text{L,R}$ in Eq.~\eqref{eq:endstates} can be combined into a nonlocal fermionic state $d_\text{M}$ with exactly zero energy.
Indeed, in the limit of infinite system size, the energy of the Majorana end modes vanishes exponentially.
(The case $w=0$ corresponds to the regime where quasi-Majorana modes are coupled pairwise with two uncoupled states at the two ends of the chain.)
In these cases, all many-body states $\ket{\psi}$ are at least doubly degenerate, since there exist at least another state with the same energy $\ket{\psi'}$, distinguished by a different value of the occupation number of the nonlocal fermionic state $n_\text{M}=0,1$.
The two degenerate states have different fermion parity, since $\{{\widetilde\gamma}_\text{L,R},P\}=0$.
We can thus define two hermitian operators
\begin{equation}\label{eq:superchargesRL}
Q_\text{L,R}=\sqrt{{\mathcal H}_\text{SUSY}}\,{\widetilde\gamma}_\text{L,R},
\end{equation}
which again satisfy the superalgebra in Eq.~\eqref{eq:hermitiansuperalgebra}.
Hence, also in this case, the Hamiltonian exhibits ${\mathcal N}=2$ SUSY with zero superpotential.
In this case, the operator $\sqrt{{\mathcal H}_\text{SUSY}}$ can be obtained by diagonalizing the Hamiltonian ${\mathcal H}_\text{eff}$ in the real space basis and taking the square roots of the diagonal elements.
For finite system sizes, the ${\mathcal N}=2$ SUSY is approximate up to exponentially small terms, i.e., it is explicitly broken by exponentially small terms which vanish as $\propto \ee^{N/\xi_\text{eff}}$.
The emergence of SUSY with supercharges $Q_\text{L,R}$ relies only on the presence of Majorana states with exponentially small energy below the particle-hole gap and therefore applies to any topological superconductor with open boundary conditions in the nontrivial phase.

\section{Spinless 1D superconductor \label{sec:pwave}}

\subsection{Minimal model}

The Majorana chain Hamiltonian in Eq.~\eqref{eq:H} can be obtained as the low-energy effective theory of a 1D topological superconductor in the presence of spatially-modulated and periodic fields.
To illustrate this idea, we first consider a simple model describing a spinless 1D topological superconductor with $p$-wave pairing and spatially-modulated chemical potential.
In the continuum, this system is described by the Hamiltonian
\begin{align}
\mathcal{H}_p=\int \dd x\Bigg[
\psi^\dag\left(\frac{p^2}{2m}-
\mu(x)
\right)\psi
+
\psi \frac{\ii \widetilde\Delta p}\hbar \psi
+\text{h.c.}
\Bigg],
\label{eq:H-pwaveC}
\end{align}
where $\psi$ is the real-space spinless fermionic field, $p=-\ii\hbar\partial_x$ is the momentum operator, $\widetilde\Delta$ the $p$-wave superconducting pairing, $m$ the effective mass, and $\mu(x)$ the space-dependent chemical potential measured with respect to the bottom of the conduction band.
If the superconducting pairing is uniform, one can always gauge away the phase of the superconducting pairing and assume $\Im \Delta=0$:
In this case, the Hamiltonian is in the BDI symmetry class~\cite{kitaev_periodic_2009,schnyder_classification_2009,ryu_topological_2010,ludwig_topological_2016} with unbroken particle-hole, time reversal, and chiral symmetries.

The Hamiltonian in Eq.~\eqref{eq:H-pwaveC} can be reduced to a discrete Hamiltonian by approximating the fields and their derivatives with finite differences on a discrete lattice $x=n a$, and by approximating the integrals with discrete sums, which yield
\begin{equation}
{\mathcal H}_p=
\sum_{n=1}^N 
\left[ (2t-\mu_n) c^\dag_n c_n - \left( t c^\dag_n c_{n+1} - \Delta c_n c_{n+1} + \text{h.c.} \right) \right],
\label{eq:H-pwaveD}
\end{equation}
where $t={\hbar^2}/{2ma^2}$ and $\Delta=\widetilde\Delta/2a$.
This is a generalization of the Kitaev chain model Hamiltonian~\cite{kitaev_unpaired_2001} with a site-dependent chemical potential $\mu_n$.

In the case of uniform fields $\mu_n=\mu$, and for small chemical potentials $|\mu|\ll 2t$, the topologically trivial and nontrivial phases are realized respectively for $\mu<0$ and $\mu>0$, and the particle-hole gap is equal to $|\mu|$.
In other words, the topological mass gap and the topological invariant are defined as $\mathcal M=-\mu$ and $\mathcal{P}=\sgn\mathcal M$, with trivial and nontrivial phases realized respectively for $\mathcal M=-\mu>0$ and $\mathcal M=-\mu<0$ ($\mathcal{P}=\pm1$).

\begin{figure}
\centering
\includegraphics[width=\columnwidth]{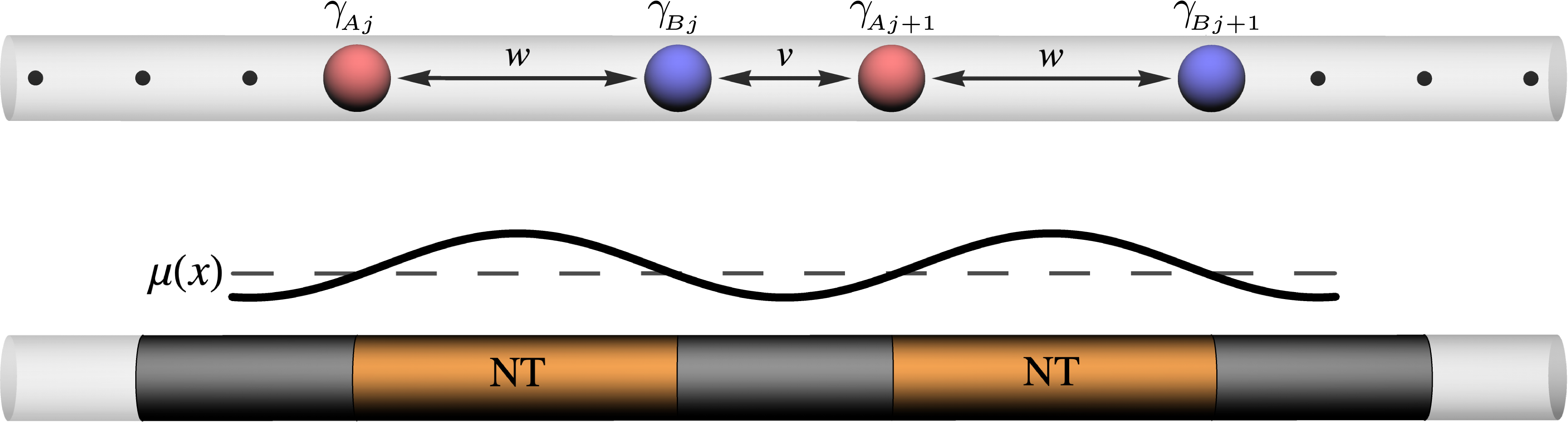}
\caption{
Spinless 1D topological superconductor with periodically-modulated chemical potential.
The variations of the chemical potential induce the presence of contiguous trivial ($\mu<0$) and nontrivial ($\mu>0$) topological segments along the wire and the emergence of localized quasi-Majorana modes at their boundaries.
}
\label{fig:pwave}
\end{figure}

If the chemical potential $\mu_n$ is not uniform, one can in principle define a local topological mass gap and topological invariant defined as $\mathcal M_n=-\mu_n$ and $\mathcal P_n=\sgn\mathcal M_n$ which may vary along the system.
If the variations of the chemical potential are smaller than the uniform term $|\delta \mu|<|\mu_0|$, the local mass gap $\mathcal M_n$ has the same sign along the system.
In particular, the system is topologically trivial ${\mathcal M}_n>0$ for $\mu_0<-|\delta \mu|$, whereas it is nontrivial ${\mathcal M}_n<0$ for $\mu_0>|\delta \mu|$.
On the other hand, if the chemical potential variations are larger than the uniform term $|\delta \mu|>|\mu_0|$, the mass gap $\mathcal M_n$ assumes alternating signs along the 1D system, as shown in \cref{fig:pwave}.
This induces alternating trivial and nontrivial segments with quasi-Majorana modes localized at the boundaries between contiguous segments.
In this case, the segments with $\mathcal M_n=-\mu_n>0$ and $\mathcal M_n=-\mu_n<0$ realize respectively topologically trivial and nontrivial phases.
In other words, these segments correspond to different signs of the \emph{local} topological mass gap $\mathcal M_n$ and different values of the topological invariant $\mathcal P_n=\sgn\mathcal M_n=\pm1$.
Consequently, quasi-Majorana modes localize at the boundaries between trivial and nontrivial segments, as shown in \cref{fig:pwave}.
The localization length of these quasi-Majorana modes is $\xi_\text{M}=\max\left(1/|\ln|z_+||,1/|\ln|z_-||\right)$, where $z_\pm$ are the solutions of the quadratic equation $(t+\Delta)z^2+(\mu-2t)+(t-\Delta)=0$.
Moreover, by requiring that the spatial variation of the chemical potential is periodic, the system becomes translational invariant:
The resulting localized quasi-Majorana modes $\gamma_{Aj}$ and $\gamma_{Bj}$ therefore realize a periodic bi-partite 1D lattice as in \cref{fig:pwave}, described by the effective low-energy Hamiltonian in Eq.~\eqref{eq:H}.
The coupling parameters $w,v$ can be calculated as the matrix elements of the Hamiltonian between quasi-Majorana modes localized at the ends of the trivial and of the nontrivial respectively, i.e., $w=-\ii\bra{\gamma_{Aj}} \mathcal{H}_{p} \ket{\gamma_{Bj}}$ and $v= -\ii\bra{\gamma_{Bj}} \mathcal{H}_{p} \ket{\gamma_{Aj+1}}$.
These overlaps are exponentially small in the distance $d$ between the quasi-Majorana modes, i.e., $|v|,|w|\propto \ee^{-d/\xi_\text{M}}$.
For $|v|=|w|$, the 1D Majorana modes become massless (gapless).
This condition coincides with the regime where the overlaps between trivial and nontrivial segments become equal:
Trivial and nontrivial segments become equivalent in the language of the low-energy effective model.
Moreover, the massless 1D Majorana modes are separated from the bulk excitations for any finite value of the chemical potential modulation $\delta\mu\neq0$.
The case $\delta\mu=\mu_0=0$ instead corresponds to the regime where the chemical potential uniformly vanishes $\mu_n\to0$, which has zero particle-hole gap.

We assume thus a periodic variation of the chemical potential with wavelength $\lambda=M a$ equal to $M$ lattice sites, which can be written in general as a sum of Fourier harmonics as
\begin{equation}\label{eq:muDFT}
\mu_n=\sum_{m=0}^{M-1} \mu_{mq} \ee^{-\ii q m n},
\quad\text{with \,}
\mu_k = \frac1M \sum_{n=0}^{M-1} \mu_n \ee^{\ii k n},
\end{equation}
where $q=2\pi/\lambda$ 
is the characteristic wavenumber of the periodic modulation.
Via Fourier transform, $c_n=(1/\sqrt{N})\sum_k \ee^{\ii k n} c_k$, the Hamiltonian in Eq.~\eqref{eq:H-pwaveD} becomes in momentum space
\begin{align}
{\mathcal H}_p&=
\sum_{k}
\sum_{m=0}^{M-1} 
h_{k+mq}\
c^\dag_{k+mq} c_{k+mq} 
+\nonumber\\&
+
\sum_{k}
\sum_{m=0}^{M-1} 
\ii \Delta \sin(k+mq) 
c_{k+mq} c_{-k+(M-m)q}
+
\text{h.c.}
+\nonumber\\&
-
\sum_{k} 
\sum_{mn=0}^{M-1}
\mu_{mq-nq}
c^\dag_{k+mq}c_{k+nq},
\end{align}
where $h_k=2t - 2t \cos{k}$, and with the momenta restricted to the first Brillouin zone $k\in[0,2\pi/\lambda]$.
Notice that the unit cell in the presence of a finite periodic chemical potential is equal to $M$ lattice sites.
Using the fermionic anticommutation relations, one can rewrite the Hamiltonian in the Bogoliubov-de~Gennes form, which reads up to a constant term
\begin{equation}\label{eq:H-pwavek}
{\mathcal H}_p=\frac12
\sum_{k}
\Psi_k^\dag
\begin{bmatrix}
\hat H(k)& \hat \Delta(k)^\dag\\
\hat \Delta(k) & -\hat H(k)^\intercal
\end{bmatrix}
\Psi_k,
\end{equation}
with
\begin{subequations}
\begin{align}
\big[ \hat H(k)\big]_{mn}&=
\diag(h_{k+mq})
-\mu_{mq-nq},
\\
\big[\hat \Delta(k)\big]_{mn}&=
2\ii\Delta\sin(k+mq)\delta_{n,M-m},
\end{align}
\end{subequations}
where $\diag(a_m)$ is the diagonal matrix and where the Kronecker delta $\delta_{nm}$ is defined modulo $M$, such that $\delta_{0M}=\delta_{M0}=1$.
In the case of a purely harmonic potential $\mu(x)=\mu_0+\delta\mu\sin(q x)$, the Hamiltonian above simplify with $\mu_{mq-nq}=\delta_{mn}\mu_0+\ii\delta\mu(\delta_{m-n,1}-\delta_{n-m,1})/2$.
Notice that the superconducting pairing couples electrons with opposite momenta $k+m q$ and $-k-mq=-k+(M-m) q$.
Moreover, one has $\hat H(k)^*=\hat H(k)^\intercal$ (hermitian) and $\hat \Delta(k)^\dag=-\hat \Delta(-k)^*$ (due to the antisymmetry of the superconducting coupling between fermions).

\begin{figure}
\centering
\includegraphics[width=\columnwidth]{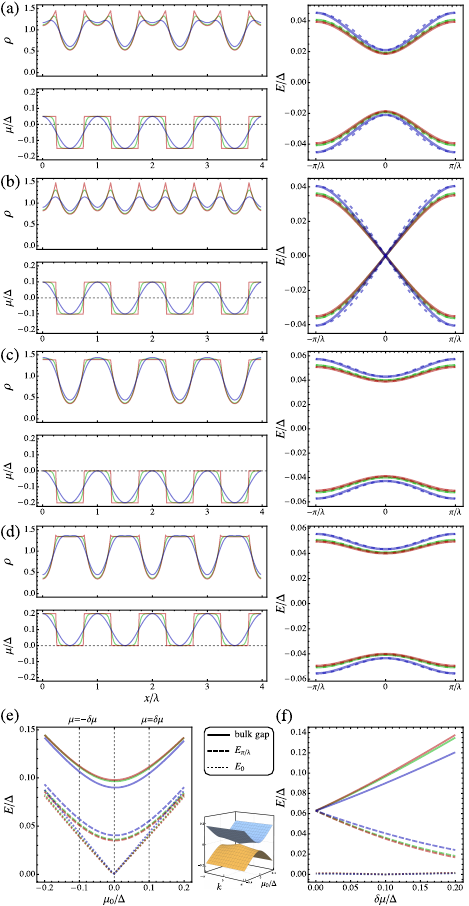}
\caption{
Probability density, chemical potential (left), and energy dispersion $E_k$ (right) of the 1D Majorana modes calculated for a $p$-wave spinless 1D superconductor with harmonic, soft-wall, and hard-wall periodic potential, with modulation amplitude $\delta\mu>0$ and $\Delta=t$.
Different rows correspond to different values of the uniform term $\mu_0=-\delta\mu/2$ (a), $\mu_0=0$ (b), $\mu_0=-\delta\mu$ (c), and $\mu_0=\delta\mu$ (d) respectively.
These choices result in massive (a), massless 1D Majorana modes (b), and gapped fermionic Andreev modes [(c) and (d)], respectively.
The peaks of the probability density correspond to the quasi-Majorana modes localized at the boundaries between the topologically trivial ($\mu<0$) and nontrivial ($\mu>0$) segments.
Dotted lines in the right panels correspond to the energy dispersion of the effective model (Majorana chain).
In the last row, energy of the 1D Majorana modes at $k=0$ (dotted lines) and $k=\pi/\lambda$ (dashed lines) compared with the bulk particle-hole gap (continuous lines) as a function of the uniform term $\mu_0$ with fixed $\delta\mu=0.1\Delta$ (e), and of the modulation amplitude $\delta\mu$ with fixed $\mu_0=0$ (f).
}
\label{fig:toymodelresults1}
\end{figure}

\subsection{Estimate of the Majorana overlaps}

The parameters $w,v$ of the effective model are defined by projecting the Hamiltonian in Eq.~\eqref{eq:H-pwaveC} onto the subspace of Majorana operators, i.e., as the matrix elements of the Hamiltonian
\begin{subequations}\label{eq:overlapestimate1}
\begin{align}
w=& -\ii \int\dd x\,\gamma_{Aj}(x)
{H}_{p}(x) 
\gamma_{Bj}(x),
\\
v=& -\ii \int\dd x\,\gamma_{Bj}(x)
{H}_{p}(x) 
\gamma_{Aj+1}(x),
\end{align}
\end{subequations}
where $H_p$ is the Hamiltonian density.
In the case of spatially-modulated chemical potential, the integrals in the equation above depend on $\mu(x)$ and on the values of $t$ and $\Delta$.
Since the Majorana wavefunctions are decaying exponentially as $\propto \ee^{-x/\xi_\text{M}}$, one has that $w\propto \ee^{-L_{AB}/\xi_\text{M}}$ and $v\propto \ee^{-L_{BA}/\xi_\text{M}}$ where $L_{AB}$ and $L_{BA}$ are the distances between contiguous quasi-Majorana modes localized at the ends of the trivial and of the nontrivial respectively, with $L_{AB}+L_{BA}=\lambda$.
Therefore, one can factor out this exponential dependence in the integrals in Eq.~\eqref{eq:overlapestimate1}.
Moreover, due to the exponential decay of the Majorana wavefuctions, the domain of integration of the integrals can be restricted only to the segments connecting contiguous quasi-Majorana states.
Thus one may write
\begin{subequations}\label{eq:overlapestimate}
\begin{align}
w&\approx \ee^{-L_{AB}/\xi_\text{M}} \int_{AB} \dd x f(\mu(x)),
\\
v&\approx \ee^{-L_{BA}/\xi_\text{M}} \int_{BA} \dd x f(\mu(x)),
\end{align}
\end{subequations}
where $f(\mu(x))$ is an unknown function of the chemical potential, and where the two domains of integration are the segments connecting contiguous quasi-Majorana states.
The two overlaps are equal in magnitude if the corresponding distances are equal, i.e., $L_{AB}= L_{BA}$ and if $|\int_{AB} \dd x f(\mu(x))|=|\int_{BA} \dd x f(\mu(x)|$.

\subsection{1D Majorana modes \label{sec:1DMajoranamodes}}

To verify the localization of the quasi-Majorana modes, we numerically diagonalize the Hamiltonian in Eq.~\eqref{eq:H-pwaveD} in real space and obtain the eigenfunctions $d=\sum_{n=1}^N \alpha_n c_n+\beta_n c_n^\dag$ of the midgap energy level (lowest positive energy) and calculate its probability density $\rho_n$ as a function of the lattice sites.
We define the probability density $\rho_n$ of the Majorana mode as the sum of the probability density of the particle and hole sectors $\rho_n=\sum_{n=1}^N |\alpha_n|^2+|\beta_n|^2$.
We then calculate the dispersion $E_k$ of the lowest energy intragap state with $E\ge0$ by diagonalizing the Hamiltonian in Eq.~\eqref{eq:H-pwavek} in momentum space for a finite system with periodic boundary conditions.
We consider different spatial dependences of the chemical potential, oscillating around the value $\mu_0$ with modulation amplitude $\delta\mu$, namely 
i) harmonic potential $\mu(x)=\mu_0+\delta\mu\cos(q x)$,
ii) periodic soft-wall potential $\mu (x+n\lambda)=\mu_0+\delta\mu\tanh{\left( \left[ \left| x- \frac{\lambda}2 \right| - \frac{\lambda}4 \right]/W\right)}$ for $0<x<\lambda$ and $n\in \mathbb{Z}$, and with wall width $W<\lambda$, and
iii) periodic hard-wall potential $\mu(x)=\mu_0+\delta\mu\sgn\cos(q x)$.

Figures~\ref{fig:toymodelresults1}(a) and \ref{fig:toymodelresults1}(b) show the probability density and the energy dispersion of the 1D Majorana modes calculated using $\mu_0=-\delta\mu/2$ (a) and $\mu_0=0$ (b) respectively.
The peaks of the density correspond to the quasi-Majorana modes localized at the boundaries between trivial ($\mu<0$) and nontrivial ($\mu>0$) segments.
As expected, the 1D Majorana modes, describing the periodic lattice of localized quasi-Majorana modes, have finite dispersion.
This is a general feature of the model in the regime $|\delta \mu|>|\mu_0|$.
In particular, we observe that for $\mu_0<0$, the overlap between quasi-Majorana modes localized at the ends of the nontrivial segments is larger than that localized at the ends of the trivial ones, whereas for $\mu_0>0$ (not shown), the overlap across the nontrivial segments is smaller.
The 1D Majorana modes are massive (gapped) and describe localized quasi-Majorana modes with finite energy and different overlaps across the trivial and nontrivial segments.
For $\mu_0=0$, the 1D Majorana modes are now massless (gapless) and describe localized quasi-Majorana modes with equal overlaps across the trivial and nontrivial segments.
The main features of the 1D Majorana modes are largely independent of the shape of the periodic modulation of the chemical potential.
However, we notice that the peaks of the probability density are much sharper in the case of hard-wall potential, whereas the peaks are smoothened out in the case of harmonic potential, with a somewhat intermediate regime in the case of soft-wall potential.
This indicates a spatial broadening of the quasi-Majorana modes in the presence of smooth variations of the potential.
This effect corresponds to a larger overlap between the wavefunctions of quasi-Majorana modes in the case of the harmonic potential, in comparison with the soft-wall and hard-wall potentials.
Larger overlaps $v,w$ translates into an enhancement of the dispersion of the 1D Majorana modes.

Figures~\ref{fig:toymodelresults1}(c) and \ref{fig:toymodelresults1}(d) show the probability density and the energy dispersion of the 1D Majorana modes calculated using $\mu_0=-\delta\mu$ (c) and $\mu_0=\delta\mu$ (d) respectively.
In these two regimes, the whole system is respectively in the trivial (c) or nontrivial (d) phases, except for a single point per unit cell (harmonic and soft-wall potential) and a segment (hard-wall potential) which realize the gapless phase with field $\mu=0$.
In these regimes, the two quasi-Majorana modes localized at the ends of the trivial and nontrivial segments merge into a single fermionic Andreev-like mode with finite energy and almost flat dispersion.
The crossover between quasi-Majorana modes and fermionic Andreev-like modes is obtained by increasing the overlap between two quasi-Majorana modes at the ends of either the trivial or nontrivial segments.

Figures~\ref{fig:toymodelresults1}(e) and \ref{fig:toymodelresults1}(f) shows the energy of the 1D Majorana modes at $k=0$ (dotted lines) and $k=\pi/\lambda$ (dashed lines) compared with the bulk particle-hole gap (continuous lines) respectively as a function of the uniform term $\mu_0$ with fixed $\delta\mu\neq0$ (e) and as a function of the modulation amplitude $\delta \mu$ with fixed $\mu_0=0$ (f).
The 1D Majorana modes coincide with the lowest positive energy level, whereas the bulk particle-hole gap is calculated as the minimum of the second-lowest positive energy level.
The 1D Majorana modes are separated from the higher-energy bulk states.
Notice that the bulk gap and the dispersion of the 1D Majorana modes are symmetric around the point $\mu_0=0$, where the mass gap of the 1D Majorana modes vanishes [compare with \cref{fig:toymodelresults1}(b)].
Moreover, by increasing the modulation amplitude of the chemical potential $\delta\mu$, the bulk gap increases, whereas the dispersion of the 1D Majorana modes decreases, approaching a flat dispersion $E\approx0$ at larger values.
Notice also that for $\delta\mu=\mu_0=0$, the energy of the 1D Majorana modes at $k=\pi/\lambda$ (dashed lines) coincides with the energy of the second-lowest energy level (continuous lines).
In this case, the particle-hole gap is closed since the dispersion of the 1D Majorana modes now connects the particle and hole sectors of the energy spectra (positive and negative energies).
This is consistent with the fact that the bulk gap of a spinless topological superconductor with uniform fields closes at $\mu=0$.

The dispersion of the 1D Majorana modes and the probability density at low-energies, which we calculated numerically, can be easily described in terms of the effective low-energy model described by the Hamiltonian in Eq.~\eqref{eq:H}.
The spatial modulation of the chemical potential generates alternating trivial and nontrivial segments, with quasi-Majorana modes localized at the boundaries between contiguous segments.
The overlaps between the wavefunctions of the quasi-Majorana modes localized at the boundaries between nontrivial and trivial segments are proportional to $w$ and $v$, respectively.
In the general case, $\mu_0\neq0$, these overlaps are different $|v|\neq|w|$ as shown in \cref{fig:toymodelresults1}(a).
Therefore, the mass gap $|v|-|w|$ of the 1D Majorana modes in Eq.~\eqref{eq:chaindispersion} is finite in this case.
In the special case $\mu_0=0$, however, the trivial and nontrivial segments are symmetrical:
Consequently, the overlaps between quasi-Majorana modes localized at the boundary of the trivial and nontrivial regions are equal $|v|=|w|$, which results in massless (gapless) 1D Majorana modes, as shown in \cref{fig:toymodelresults1}(b).
Hence, the value of the uniform term $\mu_0$ can be used to tune the value of the mass gap and to switch between massive and massless 1D Majorana modes.
Moreover, when one of the overlaps $v,w$ between two contiguous quasi-Majorana modes increases such that $|v|\gg|w|$ or $|w|\gg|v|$, the two Majoranas merge into a single fermionic state, which can be regarded as a localized Andreev mode.
We emphasize that the transition between quasi-Majorana modes and Andreev-like fermionic modes is a continuous crossover that occurs without closing the particle-hole gap of the whole system.
The fact that the mass gap of the 1D Majorana modes closes at $\mu=0$ [see \cref{fig:toymodelresults1}(d)] is also consistent with the fact that the bulk particle-hole gap must close in the uniform regime $\delta\mu\to0$.
Moreover, the fact that the dispersion of 1D Majorana modes flattens at large values of the modulation amplitude $\delta \mu$ [see \cref{fig:toymodelresults1}(e)] is a natural consequence of the fact that, in this regime, the quasi-Majorana modes become increasingly decoupled.
Consequently, their overlaps vanish $v,w\to0$, resulting in a flattening of the dispersion of the 1D Majorana modes at $E=0$.
Notice that, by increasing the modulation amplitude $\delta\mu$, the bulk gap increases, whereas the dispersion of the 1D Majorana modes decreases.

The dispersion of the 1D Majorana modes closely resembles the dispersion of the effective low-energy theory in Eq.~\eqref{eq:chaindispersion} and \cref{fig:chaindispersion}.
The effective parameter $v$ and $w$ can be obtained from the numerically calculated energy spectra as the overlap of the wavefunctions of the quasi-Majorana modes, or more straightforwardly from the energy dispersion of the energy level below the gap as $v=(E_0+E_{\pi/\lambda})/2$ and $w=(E_0-E_{\pi/\lambda})/2$ (assuming $v,w>0$).
The dotted curves in the right panels show the dispersion of the effective model (Majorana chain) in Eq.~\eqref{eq:chaindispersion}.
As we demonstrated before, the closing of the mass gap of the 1D Majorana modes is the consequence and fingerprint of SUSY\@.

\subsection{0D Majorana end modes}

\begin{figure}[t]
\centering
\includegraphics[width=.8\columnwidth]{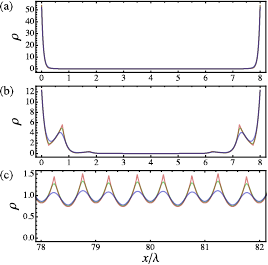}
\caption{
Probability density of the Majorana modes calculated for a $p$-wave spinless 1D superconductor with harmonic, soft-wall, and hard-wall periodic potential, with modulation amplitude $\delta\mu$ and $\Delta=t$.
Different rows correspond to different values of the uniform term $\mu_0=2\delta\mu$ (a), $\mu_0=\delta\mu/2$ (b), and $\mu_0=0$ (c) respectively.
The peaks of the probability density correspond to 
(a) Majorana end modes localized at the opposite ends of the system, with localization length 
$\xi_\text{M}\approx0.1 \lambda$,
(b) Majorana end modes localized at the opposite ends of the system, with localization length 
$\xi_\text{eff}=1/|\ln|w/v||\approx \lambda>\xi_\text{M}$, and
(c) quasi-Majorana modes localized at the boundaries between the topologically trivial and nontrivial segments in the bulk of the system.
}
\label{fig:longends}
\end{figure}

We consider now the case where the system has open boundary conditions.
As analyzed before, the system is topologically trivial or nontrivial with ${\mathcal P}={\mathcal P}_n=\pm 1$ respectively for $\mu_0>|\delta \mu|$ and $\mu_0<-|\delta \mu|$.
In the latter case, the system exhibits Majorana end modes ${\widetilde\gamma}_\text{L}$ and ${\widetilde\gamma}_\text{R}$ localized at the left and the right ends of the system.
For $\mu_0 \gg |\delta \mu|$ the localization length of these Majorana end modes coincide with the localization length $\xi_\text{M}$ of the quasi-Majorana modes $\gamma_{Aj}$ and $\gamma_{Bj}$, as shown in \cref{fig:longends} (a).

However, for $|\delta \mu|>|\mu_0|$, the local topological invariant is not uniform, and the system exhibits quasi-Majorana modes localized at the boundaries between contiguous trivial ${\mathcal M}_n>0$ and nontrivial ${\mathcal M}_n<0$ segments.
In this regime, the system is described in terms of coupled Majorana operators (Majorana chain) in the Hamiltonian in Eq.~\eqref{eq:H}.
This Majorana chain can be either trivial or nontrivial with mass gap $\mathcal{M}_\text{eff}=|w|-|v|$.
In the nontrivial case, $|v|>|w|$, the system exhibits Majorana end modes ${\widetilde\gamma}_\text{L}$, ${\widetilde\gamma}_\text{R}$ localized at the opposite ends, which are now characterized by a localization length $\xi_\text{eff}=1/|\ln|w/v||\neq \xi_\text{M}$, which depends only on the overlaps $v,w$ between contiguous quasi-Majorana modes, as shown in \cref{fig:longends} (b).
These Majorana end modes ${\widetilde\gamma}_\text{L}$, ${\widetilde\gamma}_\text{R}$ are the topologically protected end modes of the chain of quasi-Majorana modes $\gamma_{Aj}$ and $\gamma_{Bj}$ in its nontrivial state.
The transition between trivial and nontrivial phases occurs with the simultaneous closing of the particle-hole gap at $|v|=|w|$.
Therefore, for $|\delta \mu|>|\mu_0|$, the topological invariant of the system coincides with the topological invariant of the Majorana chain ${\mathcal P}=\mathcal P_\text{eff}=\sgn\mathcal M_\text{eff}$.
In the case of open boundary conditions, the massless (gapless) 1D Majorana modes are still present at $|v|=|w|$, and are localized in the bulk of the system, as shown in \cref{fig:longends} (c).
The presence of the finite boundary lifts the energy of the 1D Majorana modes.
In this case, the energy vanishes in the limit of infinite system size, as we verified numerically.

\section{Experimental realization in proximitized 1D superconductors \label{sec:swave}}

\subsection{General model}

We now describe a realistic model, i.e., a semiconducting nanowire with spin-orbit coupling (e.g., InAs, InSb) in a magnetic field and proximitized superconductivity induced by coating the wire with a conventional $s$-wave superconductor (e.g., Al, Nb) as shown in \cref{fig:swave}.
We will first introduce the Hamiltonian which describes a 1D proximitized superconductor in the presence of spatially and periodically-modulated chemical potential, magnetic fields, and superconducting pairing.
We will later discuss possible realizations of such fields in a realistic experimental setup, in particular, considering cases where either the chemical potential or the magnetic field is spatially and periodically-modulated along the wire.
In the continuum, the quantum wire is described by the Hamiltonian
\begin{align}
{\mathcal H}_s=\int \dd x\Bigg[&
\psi^\dag
\left(\frac{p^2}{2m} + \frac{\alpha}{\hbar} \sigma_y p -\mu(x)+ \mathbf{b}(x)\cdot\boldsymbol{\sigma} \right)
\psi
+\nonumber\\+&
\psi \left( \frac12 \Delta(x) \ii\sigma_y \right) \psi+\text{h.c.}
\Bigg],
\label{eq:H-swaveC}
\end{align}
where $\psi=[\psi_\up,\psi_\down]$ is the real-space electronic spinor field, $\boldsymbol\sigma=(\sigma_x,\sigma_y,\sigma_z)$ the Pauli matrices in spin space, $m$ the effective mass of the semiconducting wire, $\alpha$ the spin-orbit coupling strength, $\mu(x)$ the chemical potential measured with respect to the bottom of the conduction band, $\mathbf b(x)=({g}/2)\mu_\text{B} \mathbf{B}(x)$ the Zeeman field, and $\Delta(x)$ the superconducting pairing due to proximization.
The Hamiltonian is in the D symmetry class~\cite{kitaev_periodic_2009,schnyder_classification_2009,ryu_topological_2010,ludwig_topological_2016} with unbroken particle-hole symmetry.
We choose the cartesian axes such that the wire is parallel to the $x$-axis and the substrate surface is on the $xy$-plane (see \cref{fig:swave}).
The direction of the spin-orbit coupling term is determined by the inversion symmetry breaking at the interface between the nanowire and the substrate.
It is well known that if the magnetic field has a finite component in the $y$-direction, the bulk particle-hole gap decreases and eventually closes at a threshold angle~\cite{lin_zero-bias_2012,rex_tilting_2014,osca_effects_2014}, destroying the topological phase.
To avoid this shortcoming and maximize the bulk particle-hole gap, we will consider experimental setups where the magnetic field direction is in the $zx$-plane, i.e., perpendicular to the spin-orbit coupling term.
The superconducting pairing $\Delta(x)$ is induced via proximization~\cite{stanescu_superconducting_2013,sitthison_robustness_2014} and is comparable with the superconducting pairing of the superconducting coating, which can be orders of magnitude larger than the nominal value of the pairing in the bulk material~\cite{meservey_properties_1971}.
In the following, we will consider quantum wires either with space-dependent chemical potential induced, e.g., by periodic gating~\cite{carrad_electron_2013,burke_inas_2014}, as in \cref{fig:swave}(a), or with space-dependent magnetic (Zeeman) fields induced, e.g., by periodic arrays of nanomagnets~\cite{klinovaja_transition_2012,kjaergaard_majorana_2012,ojanen_majorana_2013,maurer_designing_2018}, or by magnetic substrates in the stripe phase~\cite{mohanta_electrical_2019,desjardins_synthetic_2019}, as illustrated in Figs.~\ref{fig:swave}(b) and \ref{fig:swave}(c).
Since both the chemical potential and the magnetic field affect the superconducting pairing, we will also consider the spatial variations of the superconducting order parameter $\Delta(x)$.
Notice that, for uniform fields, the Hamiltonian above reduces to the well-known Oreg-Lutchyn minimal model~\cite{oreg_helical_2010,lutchyn_majorana_2010}.
The Hamiltonian in Eq.~\eqref{eq:H-swaveC} can be reduced to a discrete Hamiltonian, which reads
\begin{align}
{\mathcal H}_s=&
-
\sum_{n=1}^{N-1}
c_n^\dag
\cdot 
\left(
t
+\frac\alpha{2a}\ii\sigma_y
 \right)
\cdot 
c_{n+1}
+\text{h.c.}
+\nonumber\\&
+\sum_{n=1}^N 
c_n^\dag
\cdot 
\left(
\mathbf{b}_n \cdot \boldsymbol{\sigma}+
2t-\mu_n \right)
\cdot 
c_n
+\nonumber\\&
+\sum_{n=1}^N 
c_n
\cdot 
\left(\frac12 \Delta_n \ii\sigma_y \right)
\cdot 
c_n
+\text{h.c.},
\label{eq:H-swave}
\end{align}
where $c_n^\dag=[ c^\dag_{n\up}, c^\dag_{n\down} ]$ is the spinor whose components are the fermionic operators with up and down spin on each lattice site.
 
In the case of uniform fields and assuming that the magnetic field in the $zx$-plane, the topologically nontrivial phase is realized for $b^2>\mu^2+\Delta^2$~\cite{oreg_helical_2010,lutchyn_majorana_2010}.
The topological invariant is defined as $\mathcal P=\sgn \mathcal M$, where $\mathcal M=\sqrt{\mu^2+\Delta^2}-|b|$ is the mass gap.
We notice that the lowest energy level of the Oreg-Lutchyn minimal model~\cite{oreg_helical_2010,lutchyn_majorana_2010} is unitarily equivalent to a spinless topological superconductor with a mass gap equal to $\mathcal M$ (see Ref.~\onlinecite[page 198-202]{shen_topological_2017}).
Due to the finite spin-orbit coupling, the spin of this lowest energy level is parallel to the effective magnetic field $\mathbf b+\lambda \sin{k} \hat{\mathbf{y}}$.
If the chemical potential $\mu_n$, Zeeman field $b_n$, or the superconducting pairing $\Delta_n$ are not uniform, one can consider the case where the variations of the fields are such that the local mass gap $\mathcal M_n=\sqrt{\mu_n^2+\Delta_n^2}-|b_n|$ assumes alternatively positive and negative values.
In this case, the segments with $\mathcal M>0$ and $\mathcal M<0$ realize respectively topologically trivial and nontrivial phases with a \emph{local} topological invariant $\mathcal P_n=\sgn{\mathcal M_n}=\pm1$.
Hence, quasi-Majorana modes localize at the boundaries between trivial and nontrivial segments, with localization length $\xi_\text{M}$.
The Majorana localization length is $\xi_\text{M}\approx (b_z/E_\text{SO})\alpha/\Delta$, and $\alpha/\Delta$ respectively in the weak $E_\text{SO}\ll\Delta$ and strong $E_\text{SO}\gg\Delta$ spin-orbit coupling regimes~\cite{klinovaja_composite_2012,mishmash_approaching_2016,aguado_majorana_2017}, where $E_\text{SO}=m\alpha^2/2\hbar^2$ is the energy splitting induced by the spin-orbit coupling.
Again, by requiring that the spatial variations of the fields are periodic, the system becomes translational invariant:
The Majorana modes $\gamma_{Aj}$ and $\gamma_{Bj}$ realize a periodic bi-partite 1D lattice, described by the effective low-energy Hamiltonian in Eq.~\eqref{eq:H}.
The coupling parameters $w,v$ can be again identified as the matrix elements of the Hamiltonian respectively along the nontrivial and trivial sections, i.e., $w= -\ii\bra{\gamma_{Aj}} \mathcal{H}_{s} \ket{\gamma_{Bj}}$ and $v= -\ii\bra{\gamma_{Bj}} \mathcal{H}_{s} \ket{\gamma_{Aj+1}}$.
The 1D Majorana modes become massless (gapless) for $|v|=|w|$.
In this regime, the overlaps between trivial and nontrivial segments become equal, and trivial and nontrivial segments become indistinguishable in the language of the low-energy effective model.

We consider a periodic modulation of the chemical potential, magnetic fields, and superconducting order parameter, with wavelength $\lambda=M a$ equal to $M$ lattice sites, given by Eq.~\eqref{eq:muDFT} and by
\begin{subequations}
\begin{align}\label{eq:bDFT}
b_n&=\sum_{m=0}^{M-1} b_{mq} \ee^{-\ii q m n},
\text{\, with \,}
b_k = \frac1M \sum_{n=0}^{M-1} b_n \ee^{\ii k n},
\\
\Delta_n&=\sum_{m=0}^{M-1} \Delta_{mq} \ee^{-\ii q m n},
\text{\, with \,}
\Delta_k = \frac1M \sum_{n=0}^{M-1} \Delta_n \ee^{\ii k n},
\end{align}
\end{subequations}
where $q=2\pi/\lambda$ is the wavenumber of the periodic modulation.
We assume that the wavelength $\lambda$ of the periodic modulation is much larger than the lattice parameter, so that the discrete lattice can be a good approximation of the continuum.
Via Fourier transform, the Hamiltonian in Eq.~\eqref{eq:H-swave} becomes in momentum space
\begin{align}
\mathcal H_s &=
\sum_{k}
\sum_{m=0}^{M-1} 
c^\dag_{k+mq} \cdot h_{k+mq} \cdot c_{k+mq} 
+\nonumber\\
&+
\sum_{k}
\sum_{mn=0}^{M-1} 
c_{-k+mq}
\cdot \Delta_{mq+nq} \ii \sigma_y \cdot
c_{k+nq}
+\text{h.c.}
+\nonumber\\
&+
\sum_{k} 
\sum_{mn=0}^{M-1}
c^\dag_{k+mq}
\cdot
\left(\mathbf{b}_{mq-nq}\cdot\boldsymbol\sigma -\mu_{mq-nq}\right)
\cdot
c_{k+nq},
\end{align}
where 
$h_k= 2t- 2t\cos(k)- (\alpha/a)\sin(k)\sigma_y$ and with the momenta restricted to the first Brillouin zone $k\in[0,2\pi/\lambda]$.
The length of the unit cell is equal to the wavelength $\lambda=Ma$ of the periodically-modulated fields.
Again, using the fermionic anticommutation relations, one obtains the Hamiltonian in the Bogoliubov-de~Gennes form, which reads
\begin{equation}
{\mathcal H}_s=\frac12
\sum_{k}
\Psi_k^\dag
\begin{bmatrix}
\hat H(k)& \hat \Delta^\dag\\
\hat \Delta & -\hat H(k)^\intercal
\end{bmatrix}
\Psi_k
\end{equation}
with
\begin{subequations}
\begin{align}
\big[ \hat H(k)\big]_{mn}&=
\diag(h_{k+mq})+
\mathbf{b}_{mq-nq}\cdot\boldsymbol\sigma 
-\mu_{mq-nq},
\\
\big[\hat \Delta\big]_{mn}&=
\Delta_{mq+nq} \ii \sigma_y,
\end{align}
\end{subequations}
up to a constant term.
Again, due to the antisymmetry of the superconducting coupling between fermions, one has $\hat H(k)^*=\hat H(k)^\intercal$ (hermitian) and $\hat \Delta^\dag=-\hat \Delta^*$.

The values of the parameters entering the Hamiltonians in Eqs.~\eqref{eq:H-swaveC}~and~\eqref{eq:H-swave} depend on the precise details of the specific experimental setup.
However, to do actual calculations, we take realistic values for the parameters, in agreement with previous works~\cite{mourik_signatures_2012,liu_andreev_2017}.
In particular, we take the effective mass of InSb $m=0.015\,m_\text{e}\approx\text{\SI{7500}{\eV}}/c^2$ ($m_\text{e}$ is the bare electron mass), the spin-orbit coupling parameter $\alpha=\text{\SI{1}{\eV\angstrom}}$ (which corresponds to a spin-orbit coupling length of $\approx\text{\SI{200}{\nano\metre}}$), $b/B=\text{\SI{1.5}{\milli\eV\per\tesla}}$ (which corresponds to a $g$-factor $g\approx50$), and $\Delta=\text{\SI{1}{\milli\eV}}$ at zero magnetic field.
In this regime, the spin-orbit energy splitting is $E_\text{SO}\approx\text{\SI{1}{\milli\eV}}\approx\Delta$ and the Majorana localization length is estimated to be $\xi_\text{M}\approx (b_z/E_\text{SO})\alpha/\Delta\approx\text{\SIrange{200}{300}{\nano\metre}}$.

\subsection{Characteristic wavelengths}

To realize a periodic array of spatially separated but overlapping quasi-Majorana modes along the wire, the periodic fields need to have a wavelength $\lambda$ larger than but comparable with the Majorana localization length $\xi_\text{M}$.
Indeed, the overlaps $v,w$ between contiguous quasi-Majorana modes are exponentially small in the distance $d$ between the states, i.e., $|v|,|w|\propto \ee^{-d/\xi_\text{M}}$.
Therefore, for shorter wavelengths $\lambda\lesssim\xi_\text{M}$, these overlaps $v,w$ become larger, and the groundstate and low-energy excitations of the system cannot be described in terms of spatially separated Majorana modes, as in the effective Hamiltonian in Eq.~\eqref{eq:H}.
Conversely, for very long wavelengths $\lambda\gg\xi_\text{M}$, the quasi-Majorana modes decouple, i.e., their overlaps $v,w$ become negligible, and consequently the dispersion of the 1D Majorana modes becomes nearly flat.
Moreover, the wavelength $\lambda$ must be several times smaller than the total length of the wire $L$ to avoid finite-size effects, and the wavelength $\lambda$ must be much larger than the nanowire lattice parameter $a$, so that the effects of discreteness of the lattice can be neglected.
If these two conditions are satisfied, the system can be treated effectively as an infinite system in the continuum limit.
The localization length $\xi_\text{M}$ of the quasi-Majorana modes is associated with the particle-hole minigap of the topological phase, and typically larger than the coherence length of the Cooper pairs in the proximitized superconducting wire~\cite{potter_majorana_2011,stanescu_proximity_2010,sau_nonabelian_2010}.
Also notice that the coherence length in the superconducting coating can be much larger than the coherence length in the bulk material~\cite{meservey_properties_1971}.
Typically, the localization length of Majorana modes in topological nanowires is $\xi_\text{M}\approx\text{\SIrange{100}{500}{\nano\metre}}$~\cite{stanescu_dimensional_2013}.
From these considerations, the different regimes of a 1D topological superconductor with periodically-modulated fields can be classified in terms of the three characteristic lengths $\lambda$, $\xi_\text{M}$, $a$, and $L$.
At very short wavelengths $\lambda\approx a$, the discreteness of the lattice cannot be neglected, and the system realizes a quasiperiodic regime where competition of the two periodicities, i.e., the lattice and the field periodicities, give rise to fractal energy bands (Hofstadter butterfly)~\cite{hofstadter_energy_1976} and the presence of nontrivial Andreev bound states~\cite{marra_topologically_2019}.
At intermediate wavelengths $\xi_\text{M}\gtrsim\lambda\gg a$, the overlaps between quasi-Majorana modes are so large that these cannot be treated as separate states.
This regime is realized, e.g., in the case of helical or, in general, spatially-modulated magnetic fields with periodicity $\lambda\sim\xi_\text{M}$, where the variations of the field direction induce an effective spin-orbit coupling~\cite{choy_majorana_2011,kjaergaard_majorana_2012,klinovaja_transition_2012,klinovaja_topological_2013}.
At long wavelengths and long wires $L\gg\lambda\gtrsim\xi_\text{M}\gg a$, quasi-Majorana modes are spatially separated but partially overlapping, and the system can be effectively described by the Hamiltonian in Eq.~\eqref{eq:H}.

Estimating the parameters $w,v$ of the effective model in a realistic system with spatially-modulated inhomogeneous fields is not straightforward (see, e.g., Ref.~\onlinecite{penaranda_quantifying_2018}).
Analogously to the previous case [see Eq.~\eqref{eq:overlapestimate}], these parameters can be written in terms of the local Majorana localization length as
\begin{subequations}
\begin{align}
w&\approx \ee^{-L_{AB}/\xi_\text{M}} \int_{AB} \dd x f({\mathcal M}(x)),
\\
v&\approx \ee^{-L_{BA}/\xi_\text{M}} \int_{BA} \dd x f({\mathcal M}(x)),
\end{align}
\end{subequations}
where $f({\mathcal M}(x))$ is an unknown function of the Majorana mass ${\mathcal M}(x)$, and where the two domains of integration are the segments connecting contiguous quasi-Majorana states.
In first approximation, the two overlaps are equal when the corresponding distances are equal, i.e., $L_{AB}\approx L_{BA}$.
Hence, in order to have massless 1D Majorana modes with emergent SUSY ($|v|=|w|$) one needs to satisfy the condition
\begin{equation}
L_{AB}\approx L_{BA} \approx \frac\lambda2,
\end{equation}
where $L_{AB}+ L_{BA}=\lambda$.

\subsection{1D Majorana modes realized with periodic chemical potential}

\begin{figure}
\centering
\includegraphics[width=\columnwidth]{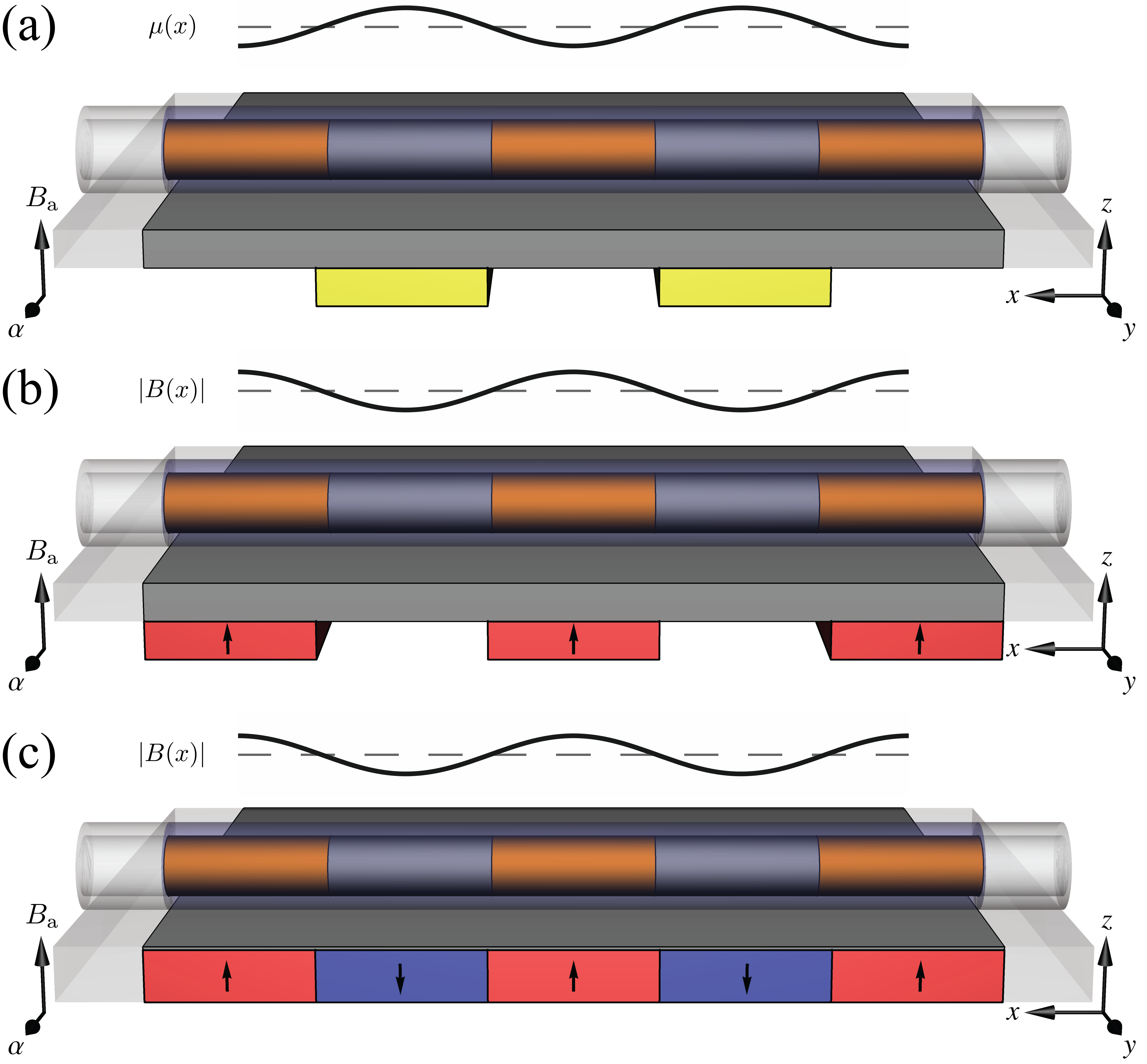}
\caption{
Section of a semiconducting nanowire with Rashba spin-orbit coupling in an external magnetic field $B_\text{a}$, and proximitized by coating with a conventional $s$-wave superconductor.
The periodic fields can be realized in different setups.
(a)
A regular array of electric gates produces a periodically-modulated chemical potential.
(b)
A regular array of nanomagnets with the same orientations or
(c) 
with alternate orientations produces a periodically-modulated magnetic field.
Magnetic stripes in a magnetic film in the stripe phase can also produce a periodically-modulated magnetic field.
The variations of the chemical potential or magnetic field induce the presence of contiguous trivial ($b^2<\mu^2+\Delta^2$) and nontrivial ($b^2>\mu^2+\Delta^2$) topological segments along the wire, and the emergence of quasi-Majorana modes localized at their boundaries.
}
\label{fig:swave}
\end{figure}

We first consider the case where the magnetic field is uniform along the wire, but the chemical potential is periodically-modulated with periodicity comparable with the Majorana localization length.
In a nanowire, this can be realized using a regular and periodic array of electrical gates at positions $x_n=n \lambda$, as illustrated in \cref{fig:swave}(a), via electron-beam patterning or wrap-gate segments~\cite{carrad_electron_2013,burke_inas_2014}.
Calculating the chemical potential $\mu(x)$ in this setup requires a self-consistent treatment of the electrostatic potentials via the Schrödinger-Poisson equation~\cite{vuik_effects_2016,woods_effective_2018}, which is beyond the scope of this work.
However, self-consistent Schrödinger-Poisson calculations have shown that the variations of the chemical potential achievable by periodic gating are quite small~\cite{woods_effective_2018,woods_enhanced_2020}.
An alternative approach is to consider 2D epitaxial semiconductor-superconductor heterostructures~\cite{shabani_two-dimensional_2016,nichele_scaling_2017,fornieri_evidence_2019} with periodical modulations of the width of the superconductor~\cite{woods_enhanced_2020}.
This system can be described as a 1D system with an effective chemical potential induced by the width modulation, which exhibits significantly larger variations~\cite{woods_enhanced_2020} compared with periodic gate setups.
For practical purposes, we thus assume a periodic dependence of the chemical potential in the form of a periodic soft-wall periodic potential
$\mu (x+n\lambda)=\mu_0+\delta\mu \tanh{\left( \left[ \left| x- \frac{\lambda}2 \right| - \frac{\lambda}4 \right]/W\right)}$ for $0<x<\lambda$ and $n\in \mathbb{Z}$ as in the previous Section.
In principle, the periodic variations of the chemical potential also affect the superconducting pairing.
To address this quantitatively, one needs to consider specific models relevant for the experimental setup and geometry, and treat the spatial variations of the chemical potential and superconducting pairing in a self-consistent way.
In general, however, one can consider the effect of the periodic modulation of the chemical potential as a small perturbation, which at the first order induces variations of the superconducting pairing proportional to the variations of the chemical potential $\delta\Delta\propto \delta\mu$.
Hence, we will take 
$\Delta(x+n\lambda)=\Delta_0+\delta\Delta \tanh{\left( \left[ \left| x- \frac{\lambda}2 \right| - \frac{\lambda}4 \right]/W\right)}$ for $0<x<\lambda$ and $n\in \mathbb{Z}$
and with $\delta\Delta=0.2\delta\mu$.

\begin{figure}
\centering
\includegraphics[width=\columnwidth]{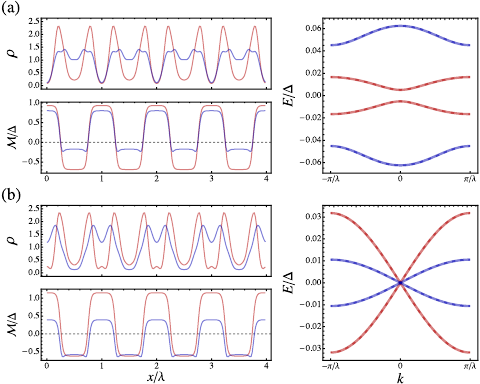}
\caption{
Probability density $\rho$ and mass gap $\mathcal M=\sqrt{\mu^2+\Delta^2}-|b|$ (left), and energy dispersion $E_k$ (right) of the 1D Majorana modes calculated for a proximitized InSb wire in the presence of a soft-wall periodic chemical potential, with modulation amplitude $\delta\mu=\Delta$ and uniform term $\mu_0=\Delta/2$.
Rows (a) and (b) correspond to different choices of the uniform term $\mu_0$ and magnetic field $b$, which result respectively in massive (gapped) and massless (gapless) 1D Majorana modes.
In the first row (a), we have $\mu_0=\Delta/2=\text{\SI{0.5}{\milli\eV}}$, $b=\sqrt{\mu^2+\Delta^2}$ (blue lines), and $b=\text{\SI{1.5}{\milli\eV}}$ ($B=\text{\SI{1}{\tesla}}$), $\mu_0=\sqrt{b^2-\Delta^2}$ (red lines), respectively.
In the second row (b), we fix $\mu_0=\Delta/2=\text{\SI{0.5}{\milli\eV}}$ (blue lines) or $B=\text{\SI{1}{\tesla}}$ (red lines), and we numerically choose $B$ and $\mu_0$, respectively, in order to obtain a gapless dispersion.
Peaks of the probability density correspond to quasi-Majorana modes localized at the boundaries between the topologically trivial ($\mathcal M>0$) and nontrivial ($\mathcal M<0$) segments.
Dotted lines correspond to the energy dispersion of the effective model (Majorana chain).
}
\label{fig:swavemodelresults1}
\end{figure}

In this setup, by tuning the magnetic field $b$ or the uniform term of the chemical potential $\mu_0$, one can realize a regime where the mass gap $\mathcal M_n=\sqrt{\mu_n^2+\Delta_n^2}-|b_n|$ assumes alternatively positive and negative values along the chain, which correspond to topologically trivial and nontrivial segments.
Figures~\ref{fig:swavemodelresults1}(a) and \ref{fig:swavemodelresults1}(b) show the probability density and the energy dispersion of the 1D Majorana modes calculated for a proximitized InSb wire with a soft-wall periodic chemical potential with amplitude $\delta\mu=\Delta$, and different choices of the uniform term $\mu_0$ and magnetic field $b$.
In the first row (a), we consider two setups with $\mu_0=\Delta/2=\text{\SI{0.5}{\milli\eV}}$, $b=\sqrt{\mu^2+\Delta^2}$ (blue lines), and $b=\text{\SI{1.5}{\milli\eV}}$ ($B=\text{\SI{1}{\tesla}}$), $\mu_0=\sqrt{b^2-\Delta^2}$ (red lines), respectively.
These choices correspond to a gapless phase with $\mathcal M=0$ in the limit of uniform field $\delta\mu\to0$.
In the second row (b), we alternatively fix $\mu_0=\Delta/2=\text{\SI{0.5}{\milli\eV}}$ (blue lines) or $B=\text{\SI{1}{\tesla}}$ (red lines), and we numerically choose the other parameter to obtain a gapless dispersion at $k=0$.
The peaks of the density correspond to quasi-Majorana modes localized at the boundaries between trivial ($\mathcal M>0$) and nontrivial ($\mathcal M<0$) segments.

\subsection{1D Majorana modes realized with periodic magnetic fields}

We now consider the case where the chemical potential is uniform along the wire, but the magnetic field is periodically-modulated and perpendicular to the spin-orbit coupling direction.
A periodic magnetic field with periodicity comparable with the Majorana localization length can be realized in at least two experimental setups.

In one setup, the periodic magnetic field is induced by a periodic arrangement of nanomagnets~\cite{klinovaja_transition_2012,kjaergaard_majorana_2012,ojanen_majorana_2013,maurer_designing_2018}.
The nanomagnets can be arranged in many different geometries~\cite{maurer_designing_2018}.
To be concrete, we will consider a regular array of nanomagnets placed below the nanowire substrate at a distance $d$ from the wire and with magnetic moments parallel to the $z$-axis and with the same orientations placed at a mutual distance $\lambda$, or with alternate orientations placed at a mutual distance $\lambda/2$, as illustrated in Figs.~\ref{fig:swave}(b) and \ref{fig:swave}(c) respectively.
Nanomagnets can be fabricated as small as hundreds of nanometers and therefore engineered into a periodic array with periodicity $\lambda$ larger but comparable with the localization length of the Majorana modes $\xi_\text{M}$.

In another possible setup, the periodic magnetic field is given by the fringing fields induced by a magnetic thin film (e.g., Co/Pt-multilayer) in the magnetic stripe phase~\cite{mohanta_electrical_2019,desjardins_synthetic_2019}.
For the sake of simplicity, we assume that a regular and periodic magnetic texture of the magnetic stripes with width $\lambda/2$, which have opposite and equal magnetic moments in the $z$-direction.
The magnetic substrate is separated from the wire by an insulating film of thickness $d$.
The magnetic stripes have a typical width~\cite{mohanta_electrical_2019,desjardins_synthetic_2019} of $\text{\SIrange{100}{200}{\nano\metre}}$, which is of the same order of magnitude of the localization length of the Majorana modes $\xi_\text{M}>\xi_\text{SC}$.
Alternatively, periodic magnetic fields can be obtained employing iron-based superconductors with incommensurate magnetic orders, i.e., spin-density waves with ordering vectors which do not belong to the reciprocal lattice of the atomic crystal structure~\cite{lee_incommensurate_2011,pratt_incommensurate_2011,wang_complex_2016,meier_hedgehog_2018,christensen_unravelling_2018,wang_broken_2019,steffensen_topological_2020}, and which may produce spatially-modulated fields with wavelength much larger than the interatomic distance.

A different approach to localize quasi-Majorana modes is to consider magnetic fields with periodic discontinuities, which may arise from topological defects of magnetic textures~\cite{control_kim_2015,fatin_wireless_2016,marra_controlling_2017}.
In particular, in magnet-superconductor hybrid nanostructures, the magnetic texture may exhibit a string of alternating and periodic disclinations, which can induce the localization of a chain of Majorana modes hybridizing into a pair of chiral modes with different velocities~\cite{majorana_rex_2020}.
In this work, however, we will only consider spatially-modulated fields which exhibit smooth spatial variations, without discontinuities and domain walls.

\begin{figure}
\centering
\includegraphics[width=\columnwidth]{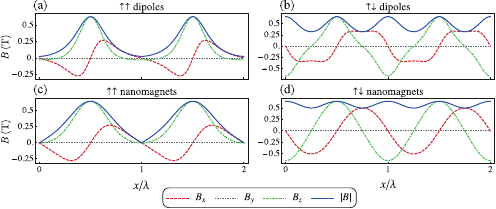}
\caption{
Magnetic field along the wire calculated in the dipole approximation for a setup with
(a)
a periodic array of equally-spaced magnetic dipoles $\mathbf{m}$, and 
(b)
a periodic array of equally spaced alternating magnetic dipoles, placed at $d=\text{\SI{250}{\nano\metre}}$ from the wire.
For reference, we use $|\mathbf{m}|=5\cdot10^{-14}\text{\si{\ampere\metre^2}}$ and $\lambda=\text{\SI{1000}{\nano\metre}}$.
Magnetic field along the wire from a finite-element method calculation for a setup with
(c)
a periodic array of nanomagnets with dimensions $\text{\SI{250}{\nano\metre}}\times\text{\SI{250}{\nano\metre}}\times\text{\SI{250}{\nano\metre}}$ and with the same orientations parallel to the $z$-axis placed at a mutual distance of $\text{\SI{750}{\nano\metre}}$ and $d=\text{\SI{250}{\nano\metre}}$ from the wire, and
(d)
a periodic array of nanomagnets with alternate orientations parallel to the $z$-axis and placed at a mutual distance of $\text{\SI{250}{\nano\metre}}$.
We consider nanomagnets with remnant magnetic field $B_\text{R}=\text{\SI{1}{\tesla}}$.
}
\label{fig:magneticfields}
\end{figure}

The periodic magnetic fields induced by a periodic array of nanomagnets or by a periodic magnetic texture can be calculated in a closed-form by using the dipole approximation.
Neglecting higher-order multipole terms, the magnetic field induced by each nanomagnet or magnetic stripe is approximately equal to the field of a magnetic dipole $\mathbf{m}_n$ in the $z$-direction, i.e., $\mathbf{B}=({\mu_{0}}/{4\pi})[{3\mathbf{r}(\mathbf{m}\cdot\mathbf{r})}/{r^{5}}-{{\mathbf{m}}}/{r^{3}}]$.
Thus the field along the wire can be written analytically as
\begin{equation}
\thinmuskip=.4\thinmuskip\medmuskip=.4\medmuskip\thickmuskip=.4\thickmuskip
\mathbf{B}(x)
=
\frac{\mu_{0}}{4\pi}
\sum_n |\mathbf{m}_n|
\frac{
3d (x-x_n) \hat{\mathbf{x}}
+
\left[ 2d^2-(x-x_n)^2 \right] \hat{\mathbf{z}}
}{
\left[d^2+(x-x_n)^2\right]^{5/2}
},
\label{eq:dipolefield}
\end{equation}
where $\mu_{0}$ is the vacuum permeability (not to be confused with the uniform component of the chemical potential used in the previous sections).
The dipole approximation is accurate for large $d$ since dipole contributions dominate at large distances.
On top of the periodically-modulated magnetic field, we consider the additional presence of a uniform external field $B_\text{a}$ in the $z$-direction, which can be used to control and tune the lengths of the topological nontrivial and trivial segments along the wire.
Figures~\ref{fig:magneticfields}(a) and \ref{fig:magneticfields}(b) show the periodic magnetic field along the wire induced respectively by a periodic arrangement of parallel magnetic dipoles 
$\mathbf{m}_n=|\mathbf{m}|\hat{\mathbf{z}}$ 
at positions $x_n=(n+1)\lambda$, and of alternating dipoles 
$\mathbf{m}_n=-(-1)^n |\mathbf{m}|\hat{\mathbf{z}}$ 
at positions $x_n=n\lambda/2$ calculated using Eq.~\eqref{eq:dipolefield}.
For large distances $d\gtrsim \lambda$, the spatial dependence of the field becomes approximately harmonic, and the variations of the field magnitude $|B|$ become smaller.
Thus, to obtain large variations of the magnetic field magnitude, one needs to minimize the distance $d$ between the wire and the magnetic dipoles.
However, in this case, the effect of higher multipole contributions cannot be neglected and, consequently, the dipole approximation in Eq.~\eqref{eq:dipolefield} cannot be accurate.
Hence, to describe a realistic system, we need to go beyond the dipole approximation.
To this aim, we calculate the magnetic field using the finite-element method with \textsc{onelab}~\cite{geuzaine_gmsh_2009}, for a periodic array of nanomagnets with magnetic moments parallel to the $z$-axis, with the same orientations and with alternate orientations.
The calculated fields are shown in Figs.~\ref{fig:magneticfields}(c) and \ref{fig:magneticfields}(d).
The spatial variations of the magnetic field induce spatial variations of the superconducting order parameter.
In order to account for the suppression of the superconducting pairing induced by the magnetic field, we assume the usual magnetic-field dependence $\Delta(x)\propto \sqrt{1-B(x)^2/B_\text{c}^2}$, where $B_\text{c}$ is the critical field of the superconducting coating, and we take $B_\text{c}=\text{\SI{1}{\tesla}}$ in numerical calculations.

\begin{figure}
\centering
\includegraphics[width=\columnwidth]{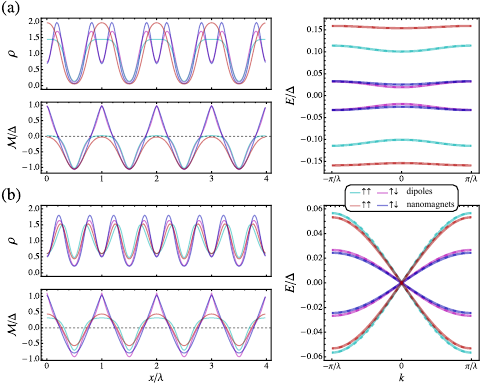}
\caption{
Probability density $\rho$ and mass gap $\mathcal M=\sqrt{\mu^2+\Delta^2}-|b|$ (left panels), and energy dispersion $E_k$ (right panels) of the 1D Majorana modes calculated for a proximitized InSb wire in the presence of periodically-modulated magnetic fields, as in \cref{fig:magneticfields}.
Rows (a) and (b) correspond to different choices of the applied magnetic field $B_\text{a}$, which result respectively in massive (gapped) and massless (gapless) 1D Majorana modes.
The peaks of the probability density correspond to quasi-Majorana modes localized at the boundaries between the topologically trivial ($\mathcal M>0$) and nontrivial ($\mathcal M<0$) segments.
Dotted lines correspond to the energy dispersion of the effective model (Majorana chain).
}
\label{fig:swavemodelresults2}
\end{figure}

Using an externally applied and uniform magnetic field $B_\text{a}$, it is possible to realize a setup where the mass gap $\mathcal M_n=\sqrt{\mu^2+\Delta^2}-|b_n|$ assumes alternatively positive and negative values along the chain.
Figures~\ref{fig:swavemodelresults2}(a) and \ref{fig:swavemodelresults2}(b) show the probability density and the energy dispersion of the 1D Majorana modes calculated for a proximitized InSb wire with periodic magnetic fields as in \cref{fig:magneticfields} and for different choices of the externally applied magnetic field.
In the first row (a), the externally applied magnetic field is chosen such that the Zeeman field is $b_\text{a}=\sqrt{\mu^2+\Delta^2}$.
This choice corresponds to a gapless phase with $\mathcal M=0$ in the limit of uniform field $b\to b_0$ (i.e., when $b_k\to 0$ for all $k\neq0$).
In the second row (b), we tune the externally applied magnetic field to obtain a gapless dispersion at $k=0$.
Again, the peaks of the density correspond to quasi-Majorana modes localized at the boundaries between trivial ($\mathcal M>0$) and nontrivial ($\mathcal M<0$) segments.
The 1D Majorana modes coincide with the superposition of the localized quasi-Majorana modes at topological phase boundaries with gapped [\cref{fig:swavemodelresults2}(a)] or gapless dispersion [\cref{fig:swavemodelresults2}(b)].
There are minimal differences between cases corresponding to magnetic fields calculated respectively in the dipole approximation and using the finite-element methods.
However, there are some noticeable differences between the configurations with magnetic moments having the same orientations or alternating orientations.
In the first case, the mass gap $\mathcal M$ becomes relatively flat close to the dipoles.
In this case, if the value of the field near the dipoles is $\mathcal M\approx 0$, the two quasi-Majorana modes at the ends of the nontrivial segments merge into a single fermionic Andreev-like mode with finite and almost flat dispersion.
The analysis of 1D Majorana modes as a function of the applied magnetic field intensity and direction can be found in the accompanying work in Ref.~\onlinecite{shortpaper}.

\begin{figure}
\centering
\includegraphics[width=.9\columnwidth]{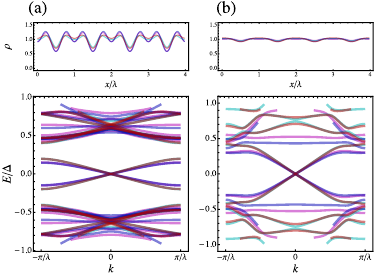}
\caption{
Probability density $\rho$ and energy dispersion $E_k$ of the 1D Majorana modes calculated as in \cref{fig:swavemodelresults2}, for different choices of the field periodicity $\lambda=\text{\SI{500}{\nano\metre}}$ (a) and $\lambda=\text{\SI{200}{\nano\metre}}$ (b).
A decrease of the field periodicity results in a decrease of the distance between quasi-Majorana modes, which finally merge into a bulk delocalized fermionic Andreev mode.
}
\label{fig:swavemodelresults3}
\end{figure}

\Cref{fig:swavemodelresults3} shows the evolution of the 1D Majorana modes and the array of localized quasi-Majorana modes as a function of the field periodicity $\lambda$.
By decreasing the field periodicity, the length of the trivial and nontrivial segments decreases and, consequently, the distance between quasi-Majorana modes at the boundaries between the trivial and nontrivial segments decreases.
This results in larger overlaps between quasi-Majorana modes, which finally merge into a bulk delocalized fermionic Andreev-like mode.
In this regime, the low energy physics cannot be described anymore as a 1D lattice of quasi-Majorana modes.

We notice that, also in the case of an $s$-wave proximitized topological superconductor in 1D, the low energy physics can be described in terms of the effective low-energy model described by the Hamiltonian in Eq.~\eqref{eq:H}.
Again, the overlaps between the wavefunctions of the quasi-Majorana modes at the ends of the nontrivial and trivial segments are proportional to $w$ and $v$, respectively.
In the general case, these overlaps are different $|v|\neq |w|$ and, consequently, the mass gap $|v|-|w|$ of the 1D Majorana modes in Figs.~\ref{fig:swavemodelresults1}(a) and \ref{fig:swavemodelresults2}(b) is finite.
However, the system can be tuned such that the overlaps are equal in magnitude $|v|=|w|$, which correspond to massless (gapless) 1D Majorana modes, as in Figs.~\ref{fig:swavemodelresults1}(b) and \ref{fig:swavemodelresults2}(b).
Moreover, when one of the overlaps $v,w$ between two contiguous quasi-Majorana modes dominates, i.e., $|v|\gg|w|$ or $|w|\gg|v|$, the two Majoranas merge into a single fermionic and localized Andreev mode:
This transition is a continuous crossover without closing the particle-hole gap of the system.
The dispersion of the 1D Majorana modes coincides with the dispersion of the effective low-energy theory in Eq.~\eqref{eq:chaindispersion}, as shown in the right panels in \cref{fig:chaindispersion} (dotted curves).
As explained before, the effective parameters $v$ and $w$ are obtained numerically from the dispersion of the energy level below the gap.
This perfect mapping is lost when the distance between quasi-Majorana modes shrinks or the bulk particle-hole gap decreases.

We will now briefly comment on the case of open boundary conditions.
If the mass gap $\mathcal M_n$ has the same sign on the whole wire, then the system realizes an insulating phase which is either topologically trivial or nontrivial, respectively for $\mathcal M_n>0$ and $\mathcal M_n<0$.
In the nontrivial phase, the wire exhibits Majorana end modes ${\widetilde\gamma}_\text{L}$, ${\widetilde\gamma}_\text{R}$ with localization length $\xi_\text{M}\approx (b_z/E_\text{SO})\alpha/\Delta$ or $\xi_\text{M}\approx \alpha/\Delta$ respectively in the weak and strong spin-orbit coupling regimes.
However, if the mass gap $\mathcal M_n$ changes its sign along the wire, the system can be described as a chain of quasi-Majorana modes, as in Eq.~\eqref{eq:H}.
The Majorana chain is either trivial ($|v|<|w|$) or nontrivial ($|v|>|w|$), with Majorana end modes ${\widetilde\gamma}_\text{L}$, ${\widetilde\gamma}_\text{R}$ having localization length $\xi_\text{eff}=1/|\ln|w/v||\neq \xi_\text{M}$ which depends only on the overlaps $v,w$ between contiguous quasi-Majorana modes.
These Majorana end modes ${\widetilde\gamma}_\text{L}$, ${\widetilde\gamma}_\text{R}$ are the topologically protected end modes Majorana chain model in its nontrivial state.
The topological invariant coincides with the topological invariant of the Majorana chain model ${\mathcal P}=\mathcal P_\text{eff}=\sgn\mathcal M_\text{eff}$.
The transition between trivial and nontrivial phases occurs when the particle-hole gap closes at $|v|=|w|$.
The massless (gapless) 1D Majorana modes are still present at $|v|=|w|$, and are localized in the bulk of the wire with open boundary conditions.

\section{Experimental signatures}

The direct signature of SUSY is the degeneracy of the many-body spectrum, i.e., the presence of supersymmetric partners with opposite fermion parities for all many-body states.
This degeneracy is present if and only if the particle-hole gap is closed.
The particle-hole gap closing can be experimentally revealed by a finite LDOS at zero-energy localized along the whole length of the wire measured, e.g., via tunneling probes placed at the bulk of the nanowire~\cite{grivnin_concomitant_2019}, or by scanning tunneling microscopy (STM) in epitaxial 1D heterostructures~\cite{shabani_two-dimensional_2016,hell_two-dimensional_2017,pientka_topological_2017,suominen_zero-energy_2017}.

Moreover, the emergence of SUSY can also be revealed by the differential conductance.
Following Ref.~\onlinecite{flensberg_tunneling_2010}, an infinite array of 0D Majorana modes with identical couplings ($v=w$) exhibits a zero-bias conductance peak given by
\begin{equation}
G=\frac{2e^2}{h} \frac{2\Gamma}{|v|+2\Gamma},
\end{equation}
where $\Gamma$ is the tunneling rate with the external lead.
This zero-bias peak corresponds to the tunneling into the 1D Majorana modes at zero energy and becomes quantized $G=2e^2/h$ in the regime where the couplings between contiguous 0D quasi-Majorana modes becomes smaller than the tunneling rate $|v|\ll\Gamma$.
For $v\neq w$ instead, the energy of the Majorana modes is lifted and, consequently, the zero-bias conductance vanishes.
Hence, for long wires or wires in a loop geometry (i.e., with closed boundary conditions), a dip-to-peak transition in the zero-bias differential conductance is a fingerprint of the SUSY at $v=w$.

We also notice that electronic interactions with the external environment may stabilize SUSY in an extended window of the parameter space.
Since contiguous 0D quasi-Majorana modes have a finite overlap, they acquire a finite charge density~\cite{lin_zero-bias_2012,ben-shach_detecting_2015}, inducing a screening charge distribution in a dielectric environment.
As demonstrated in Ref.~\onlinecite{dominguez_zero-energy_2017}, this screening charge may act back onto 0D Majorana modes, pushing their energy back to zero.
This mechanism may be generalized to the case of an array of 0D quasi-Majorana modes, as is our setup.
Hence, electronic interactions between the charge distribution of the 1D Majorana mode and an external dielectric environment may pin the gap of the 1D Majorana mode to zero, keeping the particle-hole gap closed and stabilizing SUSY\@.

\section{Conclusions \label{sec:conclusions}}

In summary, we described the realization of a dispersive (massive or massless) 1D Majorana fermion which can be constructed using a periodic lattice of 0D Majorana fermions with finite overlaps in the continuum limit, and we showed in detail the emergence of SUSY in this model.
Moreover, we described the realization of this 1D Majorana fermion model as the low-energy effective theory of 1D inhomogeneous topological superconductors, considering spinless models with $p$-wave superconducting coupling and with spatially-modulated potentials, as well as spinful models with spin-orbit coupled $s$-wave superconductivity and with spatially-modulated potentials or magnetic fields.
Our proposal can be experimentally realized by adding a single ingredient, i.e., a spatially-modulated Zeeman field or chemical potential, to a typical nanowire setup and varying the externally-applied uniform magnetic field.
Spatially-modulated Zeeman fields can be experimentally realized by periodic arrangement of nanomagnets or by employing the fringing fields induced by magnetic thin films in the magnetic stripe phase.
On the other hand, spatially-modulated chemical potentials can be realized by periodic gating in nanowires or with periodical modulations of the width of the superconductor in 2D epitaxial heterostructures.
Moreover, we established the physical requirements that allow the emergence of the 1D Majorana fermion, particularly concerning the competition of the different length scales which characterize the system.
In the massless case, these 1D Majorana modes have some remarkable properties, i.e., it exhibits conformal invariance and emergent SUSY\@.
We identified these massless Majorana modes as Goldstinos, i.e., the Nambu-Goldstone fermions associated with spontaneously broken SUSY\@.
To the best of our knowledge, this is the first example of extended SUSY with central extensions in a condensed matter system.
The experimental fingerprint of this supersymmetric massless (gapless) 1D Majorana fermions is the presence of a dip-to-peak transition in the differential conductance, a finite density of states at zero energy, and a zero-bias peak delocalized along the whole length of the wire.
This is experimentally distinguishable from the case of zero-energy peaks of 0D Majorana end modes, which are localized only at the ends of the wire.
Moreover, we emphasize that zero-energy peaks are not expected in the case of several quasi-Majorana modes at a finite distance, since the groundstate degeneracy is generally lifted by their finite overlaps, which contributes to a sizable energy shift away from zero.
A finite zero-bias peak in the presence of several quasi-Majorana modes is therefore a direct signature of the supersymmetric and massless 1D Majorana fermions.

\begin{acknowledgments}
P.~M.~thanks Sven Bjarke Gudnason, Stefan Rex, Masatoshi Sato, and Benjamin Woods for useful suggestions.
P.~M.~is supported by the Japan Science and Technology Agency (JST) of the Ministry of Education, Culture, Sports, Science and Technology (MEXT), JST CREST Grant.~No.~JPMJCR19T2, the MEXT-Supported Program for the Strategic Research Foundation at Private Universities ``Topological Science'' (Grant No.~S1511006), and Japan Society for the Promotion of Science (JSPS) Grant-in-Aid for Early-Career Scientists (Grant No.~20K14375).
D.~I.~is supported by the Financial Support of Fujukai Foundation.
M.~N.~is partially supported by the JSPS Grant-in-Aid for Scientific Research (KAKENHI) Grant No.~18H01217.
\end{acknowledgments}


\end{document}